\documentclass[aps,prb,floatfix,twocolumn,nofitinbib,hyperref=pdftex,citeautoscript]{revtex4-2}
\usepackage{epsfig}

\usepackage{amsmath,bm}
\usepackage{amssymb}
\usepackage{braket}
\usepackage{amsfonts}
\usepackage{dsfont}
\usepackage{comment,color}
\usepackage{lipsum}
\usepackage{hyperref}
\usepackage{xcolor}
\usepackage{mathdots}
\usepackage{booktabs}

\hypersetup{colorlinks=true,linkcolor=blue,anchorcolor=blue,citecolor=blue,filecolor=blue,urlcolor=blue,bookmarksnumbered=true,pdfview=FitB}

\begin{document}

\title{Floquet Majorana flat bands and emergent Cooper pair symmetries in $p$-wave magnet–superconductor heterostructure}

\author{Subhendu Kumar Patra$^1$, Gaurab Kumar Dash$^2$,
 and Manisha Thakurathi$^1$}

 \affiliation{$^{(1)}$Department of Physics, Indian Institute of Technology Hyderabad, Kandi, Sangareddy, Telangana 502285, India \\
 $^{(2)}$Department of Physics, Indian Institute of Technology Delhi, Hauz Khas, New Delhi 110016, India}

\date{\today}

\begin{abstract}
We investigate the emergence of topological superconductivity at the two-dimensional heterostructure interface between a $p$-wave magnet (pWM) and an $s$-wave superconductor. By analyzing nodal gap closings, we identify seven distinct nodal topological phases, each characterized by the presence of Majorana zero-energy flat bands and quantized zero-bias conductance peaks. We demonstrate that the effective $p$-wave nature of the system gives rise to spin-triplet pairing correlations with even-frequency, odd-parity and odd-frequency, even-parity symmetries. Notably, the introduction of inter-orbital hopping induces an exotic orbital-singlet term characterized by simultaneous odd-parity and odd-frequency. Furthermore, we explore the transition from static phases to Floquet topological regimes through periodic driving. These driven phases host both zero and $\pi$ Majorana flat bands, with transport signatures governed by the Floquet sum rule.   Most significantly, we show that periodic driving fundamentally reshapes the topological and superconducting landscape by generating multiple nodal points that support higher winding numbers and multiple Majorana flat bands, while the emergent Floquet degree of freedom doubles the number of symmetry-allowed Cooper-pair correlations. The first class of correlations is hosted by the even-Floquet sectors and has a direct counterpart in the static limit. In contrast, the second is a distinct Floquet-generated class that confines to the odd-Floquet sectors, representing a fundamentally nonequilibrium pairing channel that cannot exist in static systems. Finally, we demonstrate the robustness of these topological modes against strong disorder, confirming their potential for stable fault-tolerant applications. 
\end{abstract}

\maketitle

\section{Introduction}

Topological superconductivity \cite{RevModPhys.83.1057, doi:10.1143/JPSJ.81.011013,  PhysRevB.88.165111,PhysRevLett.111.186805,Sato_2017} has emerged as the cornerstone of modern condensed matter physics, primarily due to the potential of these systems to host Majorana zero-energy modes (MZEMs) \cite{PhysRevB.83.224511, PhysRevB.97.045415, PhysRevB.88.060504, MT1}. These boundary localized states are characterized by distinct transport signatures, such as quantized zero-bias conductance peaks, and hold significant promise for fault-tolerant topological quantum computation. Despite the theoretical appeal, the experimental realization of MZEMs is severely hindered by the scarcity of intrinsic $p$-wave superconductors in nature. Consequently, a major research thrust has shifted toward the artificial engineering of heterostructures that mimic unconventional topological superconductivity through the proximity effect. Promising candidates for such platforms involve the hybridization of conventional $s$-wave superconductors with unconventional magnets. Among these, the first-order odd-parity pWM has recently been classified as one of the unique phases of matter, defined by its coplanar but non-collinear magnetic order \cite{hellenes2024pwavemagnets,mfq8-yfsr,PhysRevB.111.165413,priessnitz2026ferroelectricpwavemagnets, li2026pwaveorbitalmagnetism}. A defining feature of the pWM is its composite time-reversal symmetry, consisting of conventional time-reversal symmetry (TRS) followed by a $\pi$-rotation in a spin space about a perpendicular axis.  Recent studies have established that the interplay between pWM and $s$-wave superconductivity alters the symmetries of Cooper pairs by inducing unconventional pairing states \cite{PhysRevB.111.144508, Fukaya_2025}. The unconventional magnetic texture of the pWM drives spin-singlet to spin-triplet conversion, which effectively causes the singlet superconductor to behave like a $p$-wave superconductor. Consequently, highly exotic pairing states emerge, such as even-frequency mixed spin-triplet with odd-parity \cite{PhysRevB.111.144508} and odd-frequency mixed spin-triplet with even-parity correlations \cite{PhysRevB.111.064502}. For that reason, this hybrid system is prone to hosting flat-band zero-energy modes in static configurations, which has been explored in \cite{flat_band, fukaya2026pwavesuperconductivityjosephsoncurrent, pal2026emergentsuperconductingphasesunconventional}. Furthermore, the intrinsic odd-parity magnetic nature of the pWM breaks current-phase symmetries and thus generates a Josephson diode effect \cite{yqsg-xdg8, fv8k-twwl, sharma2026pwavemagnetdrivenfieldfree, pal2026emergentsuperconductingphasesunconventional}. 

Beyond static heterostructures, Floquet engineering, the periodic modulation of a system's parameters, has emerged as a powerful tool for generating dynamically anomalous topological states that lack static counterparts \cite{PhysRevB.88.155133,6wnf-b5g8,69jq-rcsb}. Studies have demonstrated that periodic driving can induce Floquet Majorana Zero Energy Modes (FMZEMs) and Floquet Majorana $\pi$ Energy Modes (FMPEMs) at the edges of the Floquet Brillouin zone, protected by discrete time-translation symmetry \cite{PhysRevLett.106.220402,PhysRevB.95.155407}. Although much of the existing literature has focused on light-matter coupling in generic $p$-wave models \cite{lkf9-jgv6}, recent advancements have highlighted the utility of piecewise protocols, such as square-wave and delta-kick modulations, for their analytical tractability and efficiency in producing high-frequency topological transitions. Furthermore, research into non-equilibrium superconductivity has begun to explore how Floquet sidebands can act as a frequency-shifting mechanism, potentially converting conventional even-frequency pairings into unconventional odd-frequency correlations \cite{PhysRevB.94.094518}. 

Despite these developments, the interplay between Floquet engineering, orbital degrees of freedom, and unconventional magnetism at $p$-wave magnet/$s$-wave superconductor heterostructures remains largely unexplored, and several fundamental questions remain unanswered. Can the static topological phases of the pWM-superconductor interfaces be dynamically manipulated, enhanced, or completely generated via external periodic driving? How does the inclusion of orbital subspaces and inter-orbital coupling modify the symmetry classification of emergent Cooper pairs, particularly in the presence of unconventional magnetic order? Does the non-equilibrium regime, facilitated by Floquet sidebands, offer a mechanism to significantly broaden the manifold of permissible superconducting correlations beyond what is possible in static limits? In this paper, we provide a comprehensive investigation into the static and non-equilibrium properties of a two-dimensional heterostructure consisting of a minimal-model pWM proximitized by an $s$-wave superconductor. Our approach is divided into two primary thrusts.

We begin by establishing the topological landscape of the static system. We derive the full Bogoliubov–de Gennes (BdG) Hamiltonian in the band basis to explicitly showcase the emergence of effective $p$-wave pairing from the interplay of $p$-wave magnetism and $s$-wave superconductivity. By calculating the chiral winding numbers and the zero-bias conductance via the Blonder-Tinkham-Klapwijk (BTK) \cite{BTK_formalism} formalism, we identify seven distinct nodal topological phases. Furthermore, we provide a complete Berezinskii classification \cite{PismaZhETF.20.628, PhysRevB.45.13125} of the emergent Cooper-pair correlations, specifically examining how the inclusion of an orbital subspace and inter-orbital coupling facilitates the pairings. We extend the system into the time domain by periodically driving the chemical potential using two distinct piecewise protocols: square-wave and delta-kick modulations. We characterize the resulting FMZEMs and FMPEMs (at zero and $\pi/T$ quasi-energy, respectively) by calculating their respective winding numbers from the symmetric time-frames of the drive. The transport signatures of these driven modes are recovered using the Floquet sum rule, which accounts for the redistribution of spectral weight across multiple Floquet sidebands. A central focus of this work is the analysis of superconducting correlations in the presence of time-periodic driving. We demonstrate that the Floquet degrees of freedom act as a symmetry-shifting mechanism. We show that while the even-Floquet sectors preserve the static pairing parities, the odd-Floquet sectors host frequency-flipped counterparts. This leads to an exact doubling of the permissible Cooper-pair correlation manifold, a phenomenon we coin as "Floquet-induced symmetry doubling", providing a robust pathway for generating odd-frequency superconductivity.

Finally, we address the experimental feasibility of these phases by investigating their resilience against spatial impurities. We introduce static, random onsite disorder into the chemical potential for both the static and driven regimes. By tracking the persistence of zero and $\pi$ modes against the closing of bulk quasienergy gaps, we establish the limits of topological protection for these modes in realistic imperfect systems. The remainder of this paper is organized as follows: Section \ref{model} introduces the model Hamiltonian, and Section \ref{symmetries} analyzes its underlying symmetries to topologically characterize the MZEMs. Sections \ref{band basis}, \ref{static transport}, and \ref{static pairing} detail the emergence of effective $p$-wave pairing, steady-state transport, and the classification of emergent Cooper-pair correlations, respectively. Section \ref{floquet} extends the analysis to the non-equilibrium regime. The resulting Floquet transport signatures and the doubling of pairing correlations are discussed in Sections \ref{floquet transport} and \ref{floquet pairing}. Section \ref{disorder} evaluates the robustness of these modes against spatial disorder, followed by a concluding summary in Section \ref{conclusion}.  Details about periodic delta kick and extended Sambe space Hamiltonian are deferred to the appendices \ref{delta_kick} and \ref{extended_space}, respectively.

\begin{figure}
    \centering
    \includegraphics[width=1\linewidth]{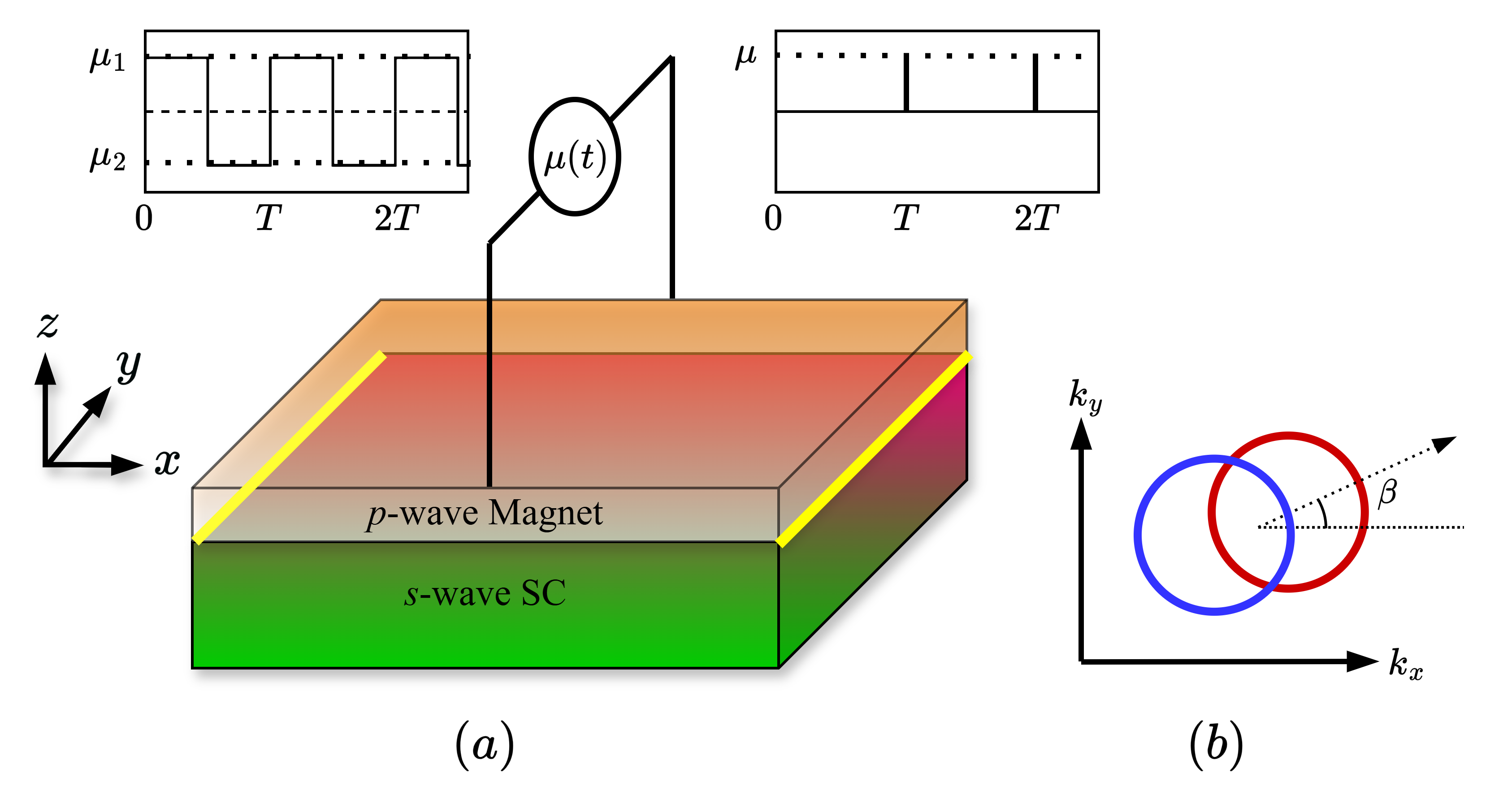}
    \caption{(a) Schematic setup of the two-dimensional heterostructure interface consisting of a two-orbital $p$-wave magnet proximitized by a conventional $s$-wave superconductor. The yellow line localized at the interface represents the Majorana Zero modes. The system is subjected to a time-periodic modulation of the chemical potential $\mu(t)$ via either square-wave or delta-kick protocols. (b) Sketch of Fermi surfaces for up (red) and down (blue) spins in pWM with lobe angle $\beta$.}
    \label{fig:schemetic}
\end{figure}

\section{The Static Model}
\label{model}
We consider a two dimensional heterostructure consisting of a pWM proximitized by an $s$-wave superconductor, as illustrated in the schematic layout of Fig. \ref{fig:schemetic}. To describe pWM, we adopt the minimal phenomenological model originally proposed in \cite{minimal_model}. The electronic degrees of freedom are described by the annhilation operator $c _{\mathbf{k},\rho,\sigma}$ corresponding to a crystal momentum $\mathbf{k}$, orbital index $\rho \in (A,B)$ and spin $\sigma \in (\uparrow,\downarrow)$. The total Hamiltonian of the heterostructure interface, expressed in the Nambu basis $\Psi_{\mathbf{k}} = [ c_{{\mathbf{k}}}, c^{\dagger}_{-{\mathbf{k}}}]^T$ with $c_{{\mathbf{k}}}= \left[ c_{{\mathbf{k}},A,\uparrow}, c_{{\mathbf{k}},A,\downarrow}, c_{{\mathbf{k}},B,\uparrow}, c_{{\mathbf{k}},B,\downarrow}\right]$, is given by:
\begin{equation}
    \mathcal{H} =  \frac{1}{2}\sum _{\mathbf{k}} \Psi_{\mathbf{k}}^{\dagger} H({\mathbf{k}}) \Psi_{\mathbf{k}},
\end{equation}
where
\begin{equation}
    H({\mathbf{k}})= 
    \begin{bmatrix}
        h_{pWM}({\mathbf{k}}) & h_{d} \\
        h_{d}^{\dagger} & -h_{pWM}^*(-{\mathbf{k}})
    \end{bmatrix}.
\end{equation}
The Hamiltonian of the pWM is expressed as:

\begin{equation}
    \begin{aligned}
    h_{pWM}({\mathbf{k}})= &-[2t (\cos k_x + \cos k_y)+\mu] \rho_0 \sigma_0 \\
    &+ [\alpha_x \sin k_x + \alpha_y \sin k_y] \rho_0 \sigma_z \\
    &+ J_{sd} \rho_z \sigma_x,
\end{aligned}
\end{equation}

where $\rho_i$ and $\sigma_i$ denote the Pauli matrices that act on the orbital and spin spaces, respectively. Here, $t$ represents the intra-orbital hopping amplitude and $\mu$ is the chemical potential. The terms $\alpha_x$ and $\alpha_y$ characterize the spin-dependent hopping strengths, which are essential for generating the non-collinear magnetic texture. These parameters scale as $\alpha_x \propto\cos(\beta)$ and $\alpha_y \propto \sin(\beta)$ where $\beta$ defines the orientation of the lobe of $p$-wave magnet relative to the $x$-axis. Consequently, for a $p_x$-wave ($p_y$-wave) magnet, $\alpha_y$ ($\alpha_x$) vanishes, whereas both terms contribute for an arbitrary lobe orientation. The parameter $J_{sd}$ represents the local exchange coupling (sd-coupling) within each orbital \cite{minimal_model}.

The proximity-induced superconductivity is introduced via the pair potential which corresponds to the $s$-wave pairing term $h_d= i d \rho_0 \sigma_y$. Together, these terms constitute a minimal effective model for investigating the topological superconductivity. 
Next, we explore the nodal phase diagram of the system for a general orientation of the lobe angle $(\beta\neq 0)$. The symmetries of the pWM Hamiltonian enforce a twofold degeneracy of the energy bands, leading to four distinct quasiparticle branches in the BdG spectrum. The eigenvalues of $H({\mathbf{k}})$ are derived as
\begin{equation}
    E_\pm^2 ({\mathbf{k}})= \mathcal{E}_{\mathbf{k}}^2 + \alpha_{\mathbf{k}}^2 + J_{sd}^2 + d^2 \pm 2\sqrt{\mathcal{E}_{\mathbf{k}}^2\alpha_{\mathbf{k}}^2 + J_{sd}^2 (\mathcal{E}_{\mathbf{k}}^2 + d^2)},
\end{equation}
where $\mathcal{E}_{\mathbf{k}}= -2t(\cos k_x+\cos k_y)-\mu$ denotes the kinetic energy contribution and $\alpha_{\mathbf{k}} =\alpha_x \sin k_x + \alpha_y \sin k_y$
 characterizes the spin-dependent hopping. The resulting BdG spectrum consists of four symmetric particle-hole branches, 
$E_+({\mathbf{k}}), E_-({\mathbf{k}}), -E_-({\mathbf{k}}), -E_+({\mathbf{k}})$, in descending order. The nodal phase boundaries separating different regimes with different numbers of nodes are identified by a closing of the bulk quasiparticle gap at zero energy that occurs when $E_-({\mathbf{k}})=0$ is satisfied, which requires
\begin{equation}
\alpha_{\mathbf{k}} =0 \quad \text{and} \quad
J_{sd}^2 = \mathcal{E}_{\mathbf{k}}^2 + d^2.
\label{eq:gap_closing_condition}
\end{equation}

We have numerically verified that the mostly the nodal points are created and annihilated at the high-symmetry points of the Brillouin zone, $ \mathbf{k} \in [\Gamma(0,0),X(0,\pm \pi),Y(\pm \pi,0),M(\pm \pi,\pm \pi)]$. Consequently, the bulk gap-closing condition can be evaluated at these high-symmetry momenta, where $\alpha_{\mathbf{k}}$ vanishes identically, yielding three critical hyperbolas that define the nodal phase boundaries in parameter space:

\begin{equation}
\begin{aligned}
    & J_{sd}^2 - \mu^2 = d^2, \quad  &&\mathbf{k} = (X,  Y), \\
    & J_{sd}^2 - (4t+\mu)^2 = d^2, \quad  &&\mathbf{k} = \Gamma, \\
    & J_{sd}^2 - (4t-\mu)^2 = d^2, \quad  &&\mathbf{k} = M. \\
\end{aligned}
\label{eq:phase_boundary}
\end{equation}
\begin{figure}
    \centering
    \includegraphics[width=1\linewidth]{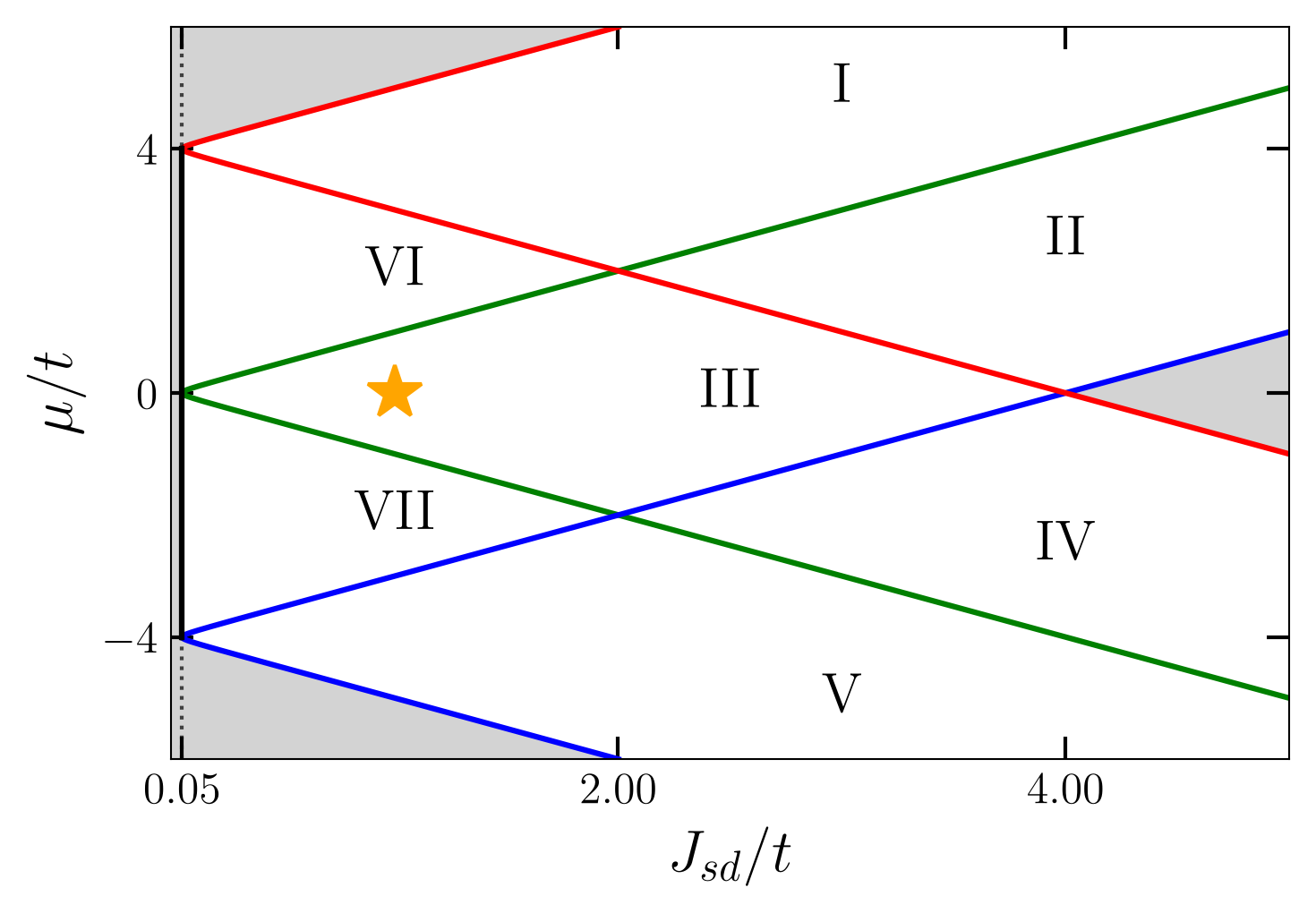}
    \caption{Illustrates the phase diagram for nodal points as a function of the exchange coupling $J_{sd}$ and the chemical potential $\mu$ for a fixed proximity-induced pair potential $d = 0.05$. The phase space is mapped by tracking the closing of the bulk energy gaps. A total of seven distinct unshaded domains are identified, while the shaded peripheral regions denote the trivial phase where the bulk is fully gapped. The transitions between these phases are represented by colors red, blue, green lines highlighting gap closing at the M point($(k_x,k_y) = (\pm\pi, \pm\pi)$), $\Gamma$ point($(k_x,k_y) = (0,0)$), X and Y points ($(k_x,k_y) = (\pm\pi,0)$ and $(0,\pm\pi)$) and black line highlighting gap closing at any other arbitrary point respectively. Phases I, II, IV, and V are characterized by two nodal points, while phases III, VI, and VII host four nodal points. A representative point of phase III is highlighted by the orange star.}
    \label{fig:phase_boundary}
\end{figure}

More generally, Eq. (\ref{eq:gap_closing_condition}) describes a continuous family of hyperbolas in the $(J_{sd}, \mu)$ parameter space for arbitrary crystal momentum $\mathbf{k}$. These hyperbolas possess vertices at $(J_{sd}, \mu)= (\pm d, -2t(\cos k_x + \cos k_y))$, whose locations span the interval $-4t<\mu<4t$. The phase boundaries given by Eq. (\ref{eq:phase_boundary}) correspond to special members of this family associated with the high-symmetry points of the Brillouin zone. The phase boundaries defined by Eq. (\ref{eq:phase_boundary}) partition the parameter space into seven distinct nodal phases, in addition to the fully gapped regions, as shown in Fig. \ref{fig:phase_boundary}. Phases I, II, IV, and V are characterized by two nodal points, while Phases III, VI, and VII host four nodal points. As the system crosses the phase boundaries highlighted by the red, green, and blue curves in Fig. \ref{fig:phase_boundary}, pairs of nodal points are either created or annihilated, signaling transitions between different nodal sectors. The red boundary corresponds to a bulk-gap closing at the M point. Consequently, crossing this boundary from the trivial region into Phase I or IV results in the creation of a pair of nodal points at M. Similarly, the blue boundary is associated with a gap closing at the $\Gamma$ point; crossing it from the trivial region to Phase II or V generates a pair of nodal points at $\Gamma$. The green boundary corresponds to simultaneous gap closings at the X and Y points such that crossing it leads to the creation or annihilation of two pairs of nodal points. In contrast, the black boundary is not associated with any high-symmetry point. Instead, crossing this boundary into Phases VI and VII generates two pairs of nodal points at generic momenta satisfying $\alpha_{\mathbf{k}}=0$, located away from conventional high-symmetry points $\Gamma$, X, Y, and M.

\section{Symmetries, Topological Invariant and MZEMs}
\label{symmetries}
The BdG Hamiltonian, by construction, preserves the particle-hole symmetry (PHS), which is expressed as:
\begin{equation}
    \mathcal{C} H({\mathbf{k}}) \mathcal{C}^{-1} = -H(-{\mathbf{k}}), \; \mathcal{C}= \tau_x \rho_0 \sigma_0\mathcal{K},
\end{equation}
where $\tau_i$'s represent the Pauli matrices acting on Nambu (particle-hole) space. Although the magnetic nature of pWM explicitly breaks the conventional time-reversal symmetry, $\mathcal{T}= i\tau_0\rho_0 \sigma_y \mathcal{K}$, the system still possesses an effective TRS. This symmetry is constructed by combining conventional TRS with a $\pi$-spin rotation about a perpendicular axis, $C_{2\perp}$, which yields an effective TRS-like operator given by
\begin{equation}
\begin{aligned}
    & \mathcal{T}'= C_{2\perp} \mathcal{T}, \text{where} ~C_{2\perp}= i \tau_z \rho_0 \sigma_z, \\
    & \mathcal{T}' H({\mathbf{k}}) \mathcal{T}'^{-1} = H(-{\mathbf{k}}). \\
\end{aligned}
\end{equation}
Furthermore, the combined action of $\mathcal{T}$ and $\mathcal{C}$ gives rise to a chiral symmetry (CS), characterized by the operator $\Gamma$, defined as
\begin{equation}
\begin{aligned}
    & \Gamma= \mathcal{T}' \mathcal{C} = -\tau_y \rho_0 \sigma_x, \\
    & \Gamma H({\mathbf{k}}) \Gamma^{-1} = -H({\mathbf{k}}).
\end{aligned}
\end{equation}
Since $(\mathcal{T}')^2 = +1$ and $(\mathcal{C})^2 = +1$, the Hamiltonian is classified in the BDI symmetry class according to the Altland–Zirnbauer scheme \cite{AZ_sym_classification}. 
The presence of chiral symmetry plays a crucial role in the topological characterization of the system, allowing its phases to be classified by an integer value winding number $W \in \mathcal{Z}$. This topological invariant provides a robust mechanism for the protection of multiple MZEMs. To characterize the zero-energy boundary states, we define a momentum-resolved one-dimensional winding number using the chiral symmetry operator $\Gamma$ for each fixed value of the transverse momentum $k_y$. This invariant can be expressed as 
\begin{equation}
    W_x(k_y) = \frac{-1}{4\pi i} \int dk_x \text{Tr} \left[ \Gamma H^{-1}({\mathbf{k}}) \partial_{k_{x}} H({\mathbf{k}}) \right].
\label{eq:winding_number}
\end{equation}
The winding number $W_x(k_y)$ directly determines the number of MZEMs localized at the boundaries for a given $k_y$ when open boundary condition (OBC) is imposed along the $x$-direction. Similarly, a winding number $W_y(k_x)$ can be defined to characterize the topological properties of a system that is finite along the $y$-direction. 
In what follows, throughout the remainder of this work we consider OBC along the $x$-axis and periodic boundary conditions along the $y$-axis. For convenience, we omit the subscript $x$ and denote the winding number as $W(k_y)$ throughout the paper unless explicitly noted otherwise.

\begin{figure}
    \centering
    \includegraphics[width=1\linewidth]{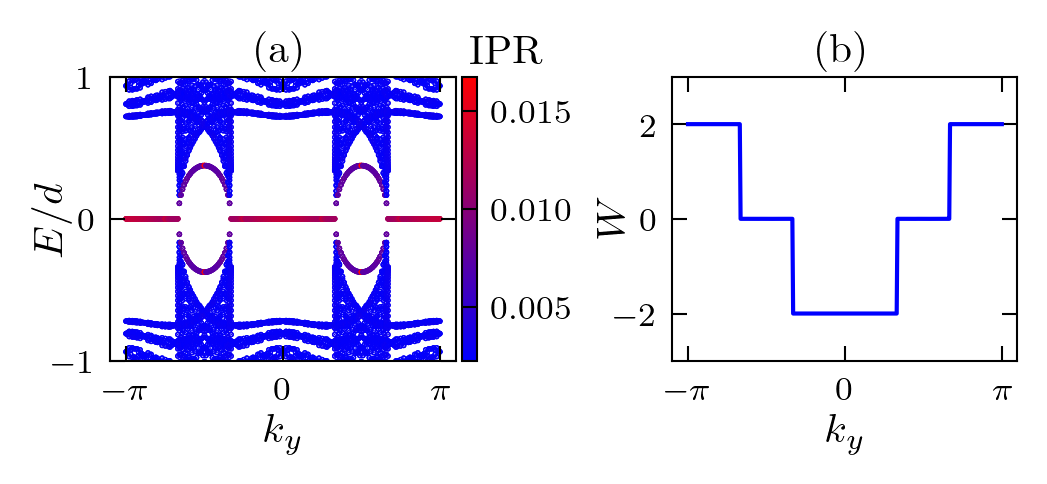}
    \caption{ (a) Showcases the eigenspectrum of the static heterostructure under a cylindrical geometry with open boundary conditions along the $x$-direction ($N_x = 500$) and periodic boundary conditions along the $y$-direction. These are color-coded with Inverse Participation Ratio (IPR), which captures the localization of the MZEMs .(b) Highlights the corresponding chiral winding number $W(k_y)$ calculated as a function of the transverse momentum to characterize the topological protection of the zero-energy manifold. The system parameters are fixed at $t = 1$, $\mu = 0$, $d = 0.05$, $\alpha_x = 1$, $\alpha_y = 0$, and $J_{sd} = 1$.}
\label{fig:energy_and_winding_static}
\end{figure}
\begin{figure}
    \centering
    \includegraphics[width=1\linewidth]{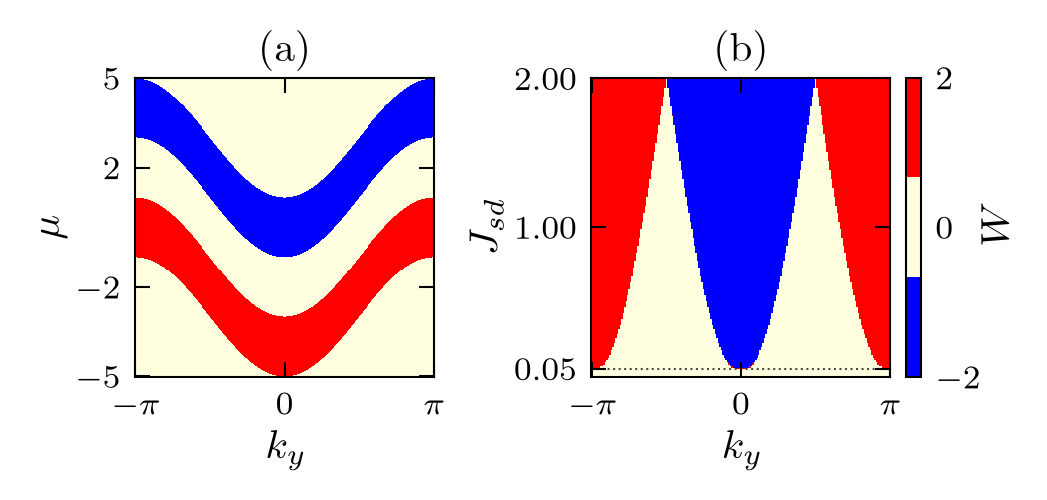}
    \caption{Topological phase diagram with varying $\mu$ and $J_{sd}$ with the transverse momentum $k_y$ along $x$-axis and with the winding number $W$ as color code. In (a) $J_{sd}$ is fixed to $1$ and in (b) $\mu$ is fixed to $0$. The remaining system parameters are the same as in Fig. \ref{fig:energy_and_winding_static}.}
\label{fig:ky_vs_mujsd}
\end{figure}

Upon imposing OBC along the $x$-direction, the nodal superconducting phase gives rise to MZEMs at the system boundaries. These edge states appear as flat zero-energy bands in the spectrum as a function of the conserved transverse momentum $k_y$ and their extent is determined by the locations of the bulk nodal points, see Fig. \ref{fig:energy_and_winding_static} (a). As a representative example, we consider a point marked by the star in Phase III of Fig. \ref{fig:phase_boundary}. In this regime, four superconducting nodes partition the $k_y$-axis into alternating topological and trivial sectors. The topological sectors host two pairs of zero-energy boundary modes, while the trivial sectors remain fully gapped. This distinction is captured by the winding number $W(k_y)$, which takes nonzero quantized values within the topological regions and vanishes elsewhere, as shown in Fig. \ref{fig:energy_and_winding_static} (b). In particular, the winding number is $\lvert W\rvert = 2$, indicating the presence of two pairs of MZEMs. The integer-valued nature of this invariant is a characteristic feature of the BDI symmetry class and establishes a direct correspondence between the bulk topology and the emergence of boundary Majorana modes in the heterostructure. The transverse momentum-resolved topological phase diagram as a function of $\mu$ is presented in Fig. \ref{fig:ky_vs_mujsd}(a). The momentum-space extent of the Majorana flat bands is found to be highly tunable with $\mu$, reflecting the strong sensitivity of the nodal topology to electrostatic gating. Furthermore, Fig. \ref{fig:ky_vs_mujsd}(b) shows the corresponding phase diagram as a function of $J_{sd}$ at $\mu=0$. We observe the critical transition at $J_{sd}/d = 1$, below which the system remains topologically trivial at all $k_y$ points. And for $J_{sd}/d > 1$, the MZEMs emerge with $W= -2~ (+2)$ localized near the zone center (boundaries). As the sd coupling is further increased, these topological regions expand in momentum space, resulting in a broadening of the Majorana flat band regime.
\begin{figure*}
    \includegraphics[width=\textwidth]{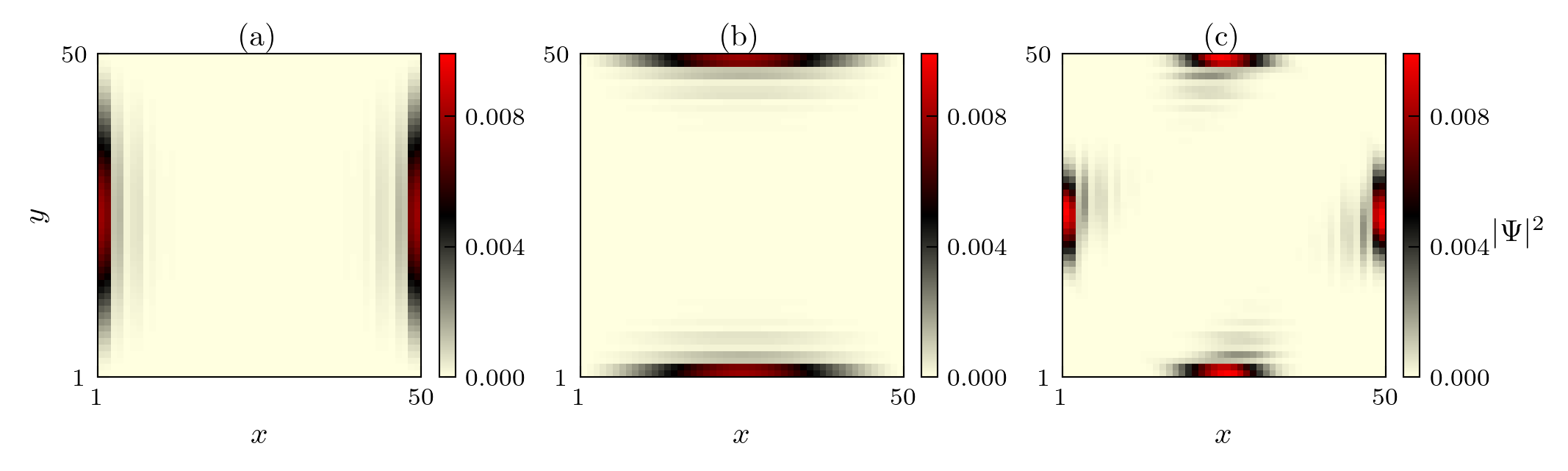}
    \caption{Real-space spatial probability distributions $|\Psi|^2$ of the Majorana zero-energy edge modes under a full OBC on a square lattice of size $N_x = N_y = 50$. Panels correspond to different configurations: (a) a pure $p_x$-wave magnet ($\alpha_x = 0, \alpha_y = 1$), (b) a pure $p_y$-wave magnet ($\alpha_x = 1, \alpha_y = 0$), and (c) a generalized $p$-wave magnet ($\alpha_x = 1, \alpha_y = 1$) with a magnetic lobe orientation angle of $\pi/4$ relative to the $x$-axis. The intensity on the color axis reflects the localized spatial profile of the wave functions, with the colorbar limits set globally to permit a direct quantitative comparison of the edge confinement across the three magnetic regimes. The system parameters are fixed at $t = 1$, $\mu = 0$, $J_{sd}/d = 2$ and $J_{sd} = 1$.}
\label{fig:OBC_xy}
\end{figure*}

Furthermore, the evolution of the Majorana flat bands across the nodal phase diagram follows directly from the creation and annihilation of bulk nodal points. The transition from the trivial gapped region (grey shaded) to the nodal phase I across the red line induces MZEMs located near $k_y = \pm\pi$. Subsequently, the crossing into Phase II via the green line extends the MZEMs to emerge near $k_y = 0$. In contrast, entering Phase V from the trivial region across the blue line yields MZEMs near $k_y = 0$, which then extend to $k_y = \pm\pi$ when entering Phase IV across the green line. In phase III, MZEMs coexist at both $k_y = 0$ and $k_y = \pm\pi$, reflecting the cumulative topological nature of the phase, whether reached from phase II or phase IV. Finally, the black lines denote direct transitions from the trivial regime into phases VI and VII. At these boundaries, the bulk gap closes at arbitrary momenta $k_y \neq (0,\pi)$, indicating a more complex nodal structure where the Majorana flat bands are delimited by non high-symmetry points. 

Imposing OBCs along both the $x$ and $y$ spatial axes reveals the localizations of MZEMs along the perimeter of a two dimensional heterostructure. The specific spatial distribution of these modes is highly sensitive to the orientation of the $p$-wave magnetic lobe ($\beta$). For the case of a $p_x$-wave magnet ($\beta = 0$), the topological gap effectively opens only relative to the $x$-direction, resulting in MZEMs that are confined exclusively to the $x= 1$ and $x= N_x$ edges. Conversely, for a $p_y$-wave magnet ($\beta = \pi/2$), the modes relocate to the $y= 1$ and $y= N_y$ boundaries. Under a general lobe orientation($\beta \neq 0,\pi/2$), both $\alpha_x$ and $\alpha_y$ terms contribute to the non-collinear magnetic texture, causing the MZEMs to manifest along all four edges of the system, as illustrated in Fig. \ref{fig:OBC_xy}. This directional tunability suggests that the $p$-wave magnetic orientation acts as a control parameter for the spatial routing of MZEMs within the heterostructure.

\section{Band Basis  Hamiltonian and effective $p$-wave pairing}
\label{band basis}

To elucidate the emergence of the MZEMs, we transform the Hamiltonian into the band basis \cite{flat_band} of the pWM. In the system Hamiltonian, the orbital Pauli matrices are restricted to $\rho_0$ and $\rho_z$, which implies a total absence of inter-orbital hybridization. As a result, the BdG Hamiltonian remains block-diagonal in the orbital basis and can be viewed as two independent superconducting subsystems, as we show below. We start by writing the normal state pWM Hamiltonian as
\begin{equation}
    h^\rho(\mathbf{k}) = \mathcal{E}_{\mathbf{k}}\sigma_0 + \alpha_{\mathbf{k}}\sigma_z + \rho J_{sd}\sigma_x.
\label{eq:normal_state_H}
\end{equation}
where $\rho= +1(-1)$ denotes the orbital $A (B)$. The resulting eigenvalues are $E_{\mathbf{k}}=\mathcal{E}_{\mathbf{k}}\pm M_{\mathbf{k}}$. The unitary matrix that diagonalizes $h^\rho (\mathbf{k})$ is defined by the rotation angle $\phi_{\mathbf{k}}$ and written as
\begin{equation}
    U_{\mathbf{k}}^\rho=     
    \begin{bmatrix}
        \cos(\frac{\phi_{\mathbf{k}}}{2}) & - \rho \sin(\frac{\phi_{\mathbf{k}}}{2}) \\
        \rho \sin(\frac{\phi_{\mathbf{k}}}{2}) & \cos(\frac{\phi_{\mathbf{k}}}{2})
    \end{bmatrix}.
\end{equation}
with  $\cos(\phi_{\mathbf{k}}) = \alpha_{\mathbf{k}}/M_{\mathbf{k}}$ and $\sin(\phi_{\mathbf{k}}) = J_{sd}/M_{\mathbf{k}}$. 
Since the spin dependent hopping $(\alpha_{\mathbf{k}})$ and the sd coupling $(J_{sd})$ together act as an effective momentum dependent magnetic field $M_{\mathbf{k}}= \sqrt {\alpha_{\mathbf{k}}^2 + J_{sd}^2}$, the diagonalized normal state pWM Hamiltonian aligns the electron spin with this local field. 
Under the full BdG transformation, $\mathcal{U}_{\mathbf{k}} ^\rho= \text{diag}\{U^\rho_{\mathbf{k}}, U^{\rho *}_{\boldsymbol{-k}} \}$, the proximity-induced $s$-wave gap $h_d = id\sigma_y$ is transformed into the band basis (represented by the pseudo-spin $\sigma '$):
\begin{equation}
\begin{aligned}
    \tilde{d}^\rho _{\mathbf{k}} =  U^{\rho \dagger}_{\mathbf{k}}(i d \sigma_y)U^{\rho *}_{\boldsymbol{-k}} & =  d^\text{od}_{\mathbf{k}}\sigma_0' + i \rho d^\text{ev}_{\mathbf{k}}\sigma_y' \\
    & = d \cos (\phi_{\mathbf{k}}) \sigma_0' + i \rho d \sin (\phi_{\mathbf{k}}) \sigma_y'
\end{aligned}
\end{equation}
This transformation reveals two distinct pairing components. While the even-parity component ($d^\text{ev}_{\mathbf{k}}  = d\sin{\phi_{\mathbf{k}}}=d J_{sd}/M_{\mathbf{k}}$) retains the $s$-wave nature, the odd-parity component ($d^\text{od}_{\mathbf{k}}  = d\cos{\phi_{\mathbf{k}}} = d\alpha_{\mathbf{k}}/M_{\mathbf{k}}$) possesses $p$-wave symmetry ($d^\text{od}_{\boldsymbol{-k}}= -d^\text{od}_{\mathbf{k}}$). This emergent odd-parity term is the fundamental driver of the nodal topological superconductivity, which results in Majorana flat bands. The full Hamiltonian in $\rho \otimes \tau \otimes \sigma'$ becomes block diagonal in the orbital sector:

\begin{equation}
    \tilde{H}(\mathbf{k})= 
    \begin{bmatrix}
        \tilde{H}^A(\mathbf{k}) & O_4 \\
        O_4 & \tilde{H}^B(\mathbf{k})
    \end{bmatrix}, 
\end{equation}
\begin{equation}
    \tilde{H}^\rho(\mathbf{k}) = 
    \begin{bmatrix}
        \mathcal{E}_{\mathbf{k}}\sigma_0' + M_{\mathbf{k}}\sigma_z' & d^\text{od}_{\mathbf{k}}\sigma_0' + i \rho d^\text{ev}_{\mathbf{k}}\sigma_y' \\
        d^\text{od}_{\mathbf{k}}\sigma_0' - i \rho d^\text{ev}_{\mathbf{k}}\sigma_y' & -\mathcal{E}_{\mathbf{k}}\sigma_0' - M_{\mathbf{k}}\sigma_z'
    \end{bmatrix}.
\end{equation}

One can clearly see the appearance of two pairs of MZEMs originating from the orbital structure of the Hamiltonian.  Within this decoupled regime, the total topological invariant is the sum of the winding numbers for each individual orbital sector, i.e., $W = W_A + W_B$. When the system enters a topological phase, each orbital sector independently undergoes a transition and hosts a single pair of MZEMs at the boundaries. The resulting $\lvert W\rvert = 2$ quantization therefore reflects the orbital multiplicity of the underlying topological phase.

\section{Transport Properties of MZEM}
\label{static transport}
To identify experimentally accessible signatures of the MZEMs, we analyze the charge transport through a normal metal–superconductor (NS) junction. Specifically, we consider a single lead setup with point contact, in which a normal metallic lead is attached to the left edge ($x=1$) of the heterostructure. The system remains translationally invariant along the $y$-direction, allowing the transport response to be resolved as a function of the conserved transverse momentum $k_y$.

The coupling between the lead and the heterostructure is described by the matrix $\mathbf{V}$. Throughout this work, we employ the wide-band limit (WBL), where the lead density of states $\rho_{l}$ is taken to be energy independent. Under this approximation, the retarded Green's function of the isolated lead takes a simple form $\mathbf{g}^R=-i\pi \rho_{l} \mathbf{I}$. Integrating out the lead degrees of freedom yields the self-energy contribution to the device region as

\begin{equation}
    \mathbf{\Sigma}= \mathbf{V}^\dagger \mathbf{g}^R \mathbf{V}.
\end{equation}
The broadening matrix is then defined as $\mathbf{\Gamma}= i(\mathbf{\Sigma}-\mathbf{\Sigma}^{\dagger})$. The retarded Green's function of the system in the presence of the lead is given by:
\begin{equation}
    \mathbf{G}^R(E)= [(E + i \eta)\mathbf{I} - \mathbf{H} - \mathbf{\Sigma}]^{-1}.
\end{equation}

Using the Fisher-Lee approach \cite{Fisher-Lee, Datta_1995}, the scattering matrix is constructed from the retarded Green's function and the broadening matrix as
\begin{equation}
    \mathbf{S}_{\alpha\beta}(E) =  \delta_{\alpha\beta} \mathbf{I} - i \sqrt{\mathbf{\Gamma}_{\alpha}} \mathbf{G}^R (E) \sqrt{\mathbf{\Gamma}_{\beta}},
\end{equation}
where $\alpha,\beta \in \{L,R\}$ denote the leads. In our single-lead setup, we focus only on the left reflection block $S_{LL}(E)$, which in the Nambu basis takes the form:
\begin{equation}
    \textbf{S}_{LL}(E) = \textbf{r}(E) = 
    \begin{bmatrix}
        \textbf{r}_{ee}(E) & \textbf{r}_{eh}(E) \\
        \textbf{r}_{he}(E) & \textbf{r}_{hh}(E)
    \end{bmatrix}.
\end{equation}

Now, since the lead is coupled solely to the first site with coupling strength $t_c$, the coupling matrix is zero everywhere except the contact site. For this first contact block, it takes the form $\mathbf{V}_c(=-t_c \tau_z\rho_0\sigma_0)$ (subscript c represents contact site block throughout this section). As a consequence, the self-energy and broadening matrices become constant matrices with a non-zero contact block only.  
\begin{equation}
    \mathbf{\Sigma}_c = -i\pi t_c^2\rho_{l}\mathbf{I}_c, \quad \mathbf{\Gamma}_c = \Gamma_0 \mathbf{I}_c.
\end{equation}
with $\Gamma_0 = 2 \pi t_c^2 \rho_l$. At the contact site, the non-zero local reflection block evaluates to
\begin{equation}
    \textbf{r}_c(E) = \mathbf{I}_c - i \Gamma_0 \mathbf{G}_c^R.
    \label{rc}
\end{equation}

The differential conductance $\sigma(V)$ is then calculated via the BTK formula \cite{BTK_formalism}:

\begin{equation}
    \sigma(V) = \frac{e^2}{h} \left[ N - R_{ee}(V) + R_{he}(V) \right],
\end{equation}
where $R_{ee} = \text{Tr}[\mathbf{r}_{ee}^\dagger \mathbf{r}_{ee}]$ and $R_{he} = \text{Tr}[\mathbf{r}_{he}^\dagger \mathbf{r}_{he}]$ denote the probabilities of normal and local Andreev reflection, respectively, and $N$ is the number of propagating electron channels in the normal lead. In the topological regime, MZEMs mediate resonant Andreev reflection, producing quantized zero-bias conductance peaks.

\begin{figure}
    \centering
    \includegraphics[width=1\linewidth]{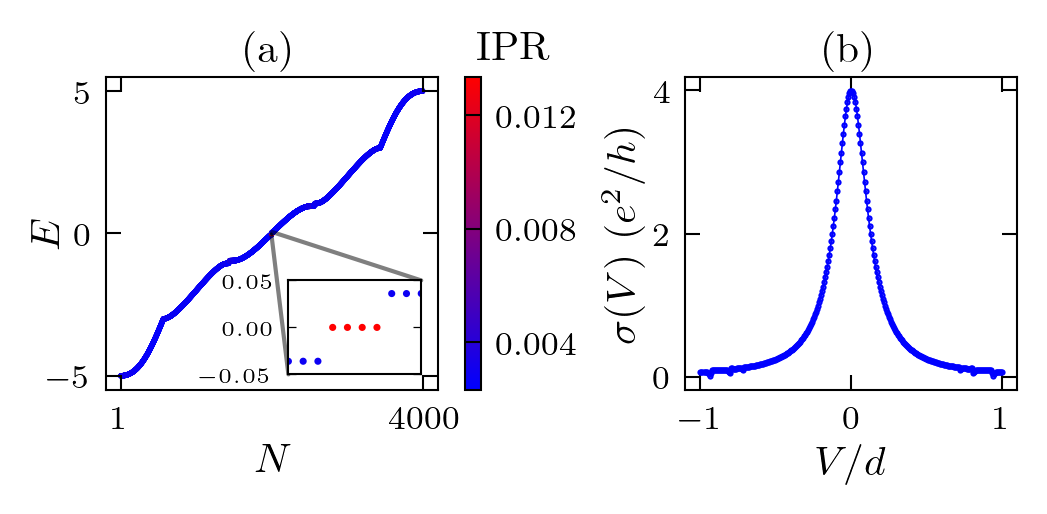}
    \caption{Depicts transport signatures and zero-bias conductance quantization for a fixed transverse momentum $k_y = 0$ under a cylindrical geometry with open boundary conditions along the $x$-direction ($N_x = 500$). (a) Shows the quasiparticle energy spectrum highlighting the isolated, fourfold-degenerate Majorana zero-energy states pinned strictly at zero energy within the bulk superconducting gap. (b) Highlights differential conductance $\sigma(V)$ as a function of the bias voltage $V$, calculated within the BTK formalism. The clear quantization of the zero-bias conductance peak ($V = 0$) to a value of $4e^2/h$ directly reflects the topological contribution of the two pairs of degenerate Majorana modes. The remaining system parameters are the same as in Fig. \ref{fig:energy_and_winding_static} with the interface barrier strength of $\Gamma_0 = 0.2$.}
\label{fig:transport_static}
\end{figure}

As a representative example, we consider $k_y = 0$ in Phase III ($W = -2$). The corresponding energy spectrum exhibits four zero-energy states in the finite strip geometry, with two modes localized at each boundary (refer Fig. \ref{fig:transport_static} (a)). Our transport calculations show that these modes manifest themselves as a quantized zero-bias conductance peak of $4e^2/h$ (see Fig. \ref{fig:transport_static}(b)).  This quantization originates from two independent Majorana-induced perfect Andreev reflection channels, each contributing $2e^2/h$, and therefore constitutes a clear transport signature of the $W = -2$ topological phase.

\section{Emergent Cooper Pair Correlations}
\label{static pairing}
In a superconductor, the Cooper pair amplitudes are characterized by the anomalous Green's function, defined as:
\begin{equation}
    F_{\rho_1,\rho_2; \sigma_1,\sigma_2}(\mathbf{k}_1,\mathbf{k}_2;t_1,t_2)= -i\langle\mathbb{T} c_{\rho_1,\sigma_1}(\mathbf{k}_1;t_1)c_{\rho_2,\sigma_2}(\mathbf{k}_2;t_2)\rangle.
\end{equation}

Due to the fermionic nature of electrons, the total pairing amplitude must satisfy the generalized anti-symmetry condition under the exchange of orbital, spin, momentum, and time (frequency) degrees of freedom:
\begin{equation}
F_{\rho_1,\rho_2;\sigma_1,\sigma_2}(\mathbf{k}_1,\mathbf{k}_2;t_1,t_2)= -F_{\rho_2,\rho_1;\sigma_2,\sigma_1}(\mathbf{k}_2,\mathbf{k}_1;t_2,t_1).
\label{F}
\end{equation} 

To characterize the superconducting Cooper-pair correlations induced in our heterostructure, we evaluate the Nambu-Gorkov Green's function and extract the Anomalous component $(F= G_{eh})$. In the presence of orbital and spin degrees of freedom, $F$ is a $4 \times 4$ matrix. The proximity induced $s$-wave pairing potential is given by intra-orbital spin-singlet, $h_d= i d  \rho_0 \sigma_y$. As noted earlier, the pWM Hamiltonian contains only intra-orbital couplings. Consequently, the full $8 \times 8$ BdG Hamiltonian decouples into two independent sectors in orbital space. These two sectors are identical except for the sign of the term $J_{sd}$, which corresponds to $\rho = +1 (-1)$ for orbital A  (B). The effective normal-state Hamiltonian describing each orbital sector is written in Eq.(\ref{eq:normal_state_H}). The corresponding Nambu-Gorkov Green's function for these orbital blocks are given by
\begin{equation}
\begin{aligned}
    \mathbf{G}^\rho(z,\mathbf{k}) & = 
    \begin{bmatrix}
       G_0^\rho(z,{\mathbf{k}}) & F^\rho(z,{\mathbf{k}}) \\
        F^\rho {^\dagger} (z,{\mathbf{k}})  & \tilde{G}_0 ^\rho(z,{\mathbf{k}})
    \end{bmatrix} \\
    & =
    \begin{bmatrix}
       S_{\mathbf{k}} -\rho J_{sd}\sigma_x & -i d \sigma_y \\
        i d \sigma_y & \tilde{S}_{\mathbf{k}} + \rho J_{sd}\sigma_x
    \end{bmatrix} ^{-1}.
\end{aligned}
\end{equation}
 where $S_{\mathbf{k}} = (z-\mathcal{E}_{\mathbf{k}})\sigma_0 - \alpha_{\mathbf{k}}\sigma_z$ and $\tilde{S}_{\mathbf{k}} = (z+\mathcal{E}_{\mathbf{k}})\sigma_0 - \alpha_{\mathbf{k}}\sigma_z$ with $z= \omega+i\eta$, $\eta \rightarrow 0$. Utilizing the block-matrix inversion identity for a matrix $M =  \begin{bmatrix}
    M_{11} & M_{12} \\
    M_{21} & M_{22}
\end{bmatrix}$ and its inverse $N =  \begin{bmatrix}
    N_{11} & N_{12} \\
    N_{21} & N_{22}
\end{bmatrix}$, the upper-right anomalous block is obtained as, $N_{12}= [M_{21} - M_{22}M_{12}^{-1}M_{11}]^{-1}$. Following this analytical inversion, the anomalous Green's functions $F^\pm(z, {\mathbf{k}})$ can be expressed as:
\begin{equation}
\begin{aligned}
    F^\pm(z, {\mathbf{k}}) =  & \frac{d}{\Delta} [ \mp 2 \alpha_{\mathbf{k}} J_{sd}\sigma_0 + 2 \alpha_{\mathbf{k}} \mathcal{E}_{\mathbf{k}}\sigma_x \\
    & + i (z^2 - \mathcal{E}_{\mathbf{k}}^2 - d^2 - \alpha_{\mathbf{k}}^2 + J_{sd}^2)\sigma_y \mp 2 z J_{sd} \sigma_z ],
\end{aligned}
\end{equation}
where the denominator is $\Delta = -(z^2 - \mathcal{E}_{\mathbf{k}}^2 - d^2 - \alpha_{\mathbf{k}} ^2 + J_{sd}^2)^2 + 4\mathcal{E}_{\mathbf{k}}^2 \alpha_{\mathbf{k}}^2 + 4J_{sd}^2(z^2 - \alpha_{\mathbf{k}}^2)$. Since $\Delta$ contains only even powers of $z$ and $\mathbf{k}$, the symmetry of the pairing amplitudes is entirely determined by the numerators. The full $4 \times 4$ anomalous matrix is reconstructed as $F = \rho_0\otimes \frac{F^+ + F^-}{2}  + \rho_z \otimes \frac{F^+ - F^-}{2}$. In the rotated spin and orbital basis, the general pairing matrix is decomposed as:
\begin{equation}
\begin{aligned}
    F(z,{\mathbf{k}}) &= [f_{ss}^{os} \rho_0 \sigma_0 + f_{st1}^{os}\rho_0 \sigma_x + f_{st2}^{os}\rho_0 \sigma_y + f_{st3}^{os}\rho_0 \sigma_z ] \\
    & + [f_{ss}^{ot1} \rho_x \sigma_0 + f_{st1}^{ot1}\rho_x \sigma_x + f_{st2}^{ot1}\rho_x \sigma_y + f_{st3}^{ot1} \rho_x \sigma_z ] \\
    & + [f_{ss}^{ot2} \rho_y \sigma_0 + f_{st1}^{ot2}\rho_y \sigma_x + f_{st2}^{ot2}\rho_y \sigma_y + f_{st3}^{ot2} \rho_y \sigma_z ] \\
    & + [f_{ss}^{ot3} \rho_z \sigma_0 + f_{st1}^{ot3}\rho_z \sigma_x + f_{st2}^{ot3}\rho_z \sigma_y + f_{st3}^{ot3} \rho_z \sigma_z ] \\
    & .(i\rho_y \otimes i\sigma_y),
\end{aligned}
\end{equation}
where the subscript $os$ and $ss$ denote the orbital and spin singlet and $(ot1,ot2,ot3)$ and $(st1,st2,st3)$ denote the orbital and spin triplet channels. The presence of orbital degrees of freedom expands the basis to $16$ possible symmetry channels. In this configuration, only four channels survive:
\begin{equation}
\begin{aligned}
    & f_{ss}^{ot2} = \frac{d}{\Delta}i(z^2 - \mathcal{E}_{\mathbf{k}}^2 - d^2 - \alpha_{\mathbf{k}}^2 + J_{sd}^2), \\ 
    & f_{st3}^{ot2} = \frac{d}{\Delta} 2i \mathcal{E}_{\mathbf{k}} \alpha_{\mathbf{k}}, \\
    & f_{st2}^{ot1} = \frac{d}{\Delta} 2i \alpha_{\mathbf{k}} J_{sd}, \\
    & f_{st1}^{ot1} = \frac{d}{\Delta} 2zJ_{sd}.
\end{aligned}
\label{eq:static_pairing_amp}
\end{equation}

The primary term $f_{ss}^{ot2}$ represents the native even-frequency, even-parity spin-singlet pairing inherited from the parent superconductor. The remaining terms $f_{st3}^{ot2}$, $f_{st2}^{ot1}$ and $f_{st1}^{ot1}$ correspond to emergent spin-triplet correlations induced by the magnetic texture of the pWM interface. The spin triplet terms, $f_{st3}^{ot2}$ and $f_{st2}^{ot1}$, exhibit an explicit momentum dependence through $\alpha_{\mathbf{k}}$. Since $\alpha_{\mathbf{k}}$ characterizes the magnetic order of the normal-state Hamiltonian, these terms emerge as a direct consequence of the odd-parity magnetic nature of the system. Notably, these amplitudes are frequency-independent; thus, the required fermionic antisymmetry is satisfied by their odd-parity spatial symmetry. In contrast, the final spin-triplet term, $f_{st1}^{ot1}$, lacks momentum-dependent factors in the numerator, but is directly proportional to the frequency $z$. Here, the antisymmetry of the Cooper pair is preserved by the odd-frequency dependence in conjunction with even spatial parity. The emergence of the $f_{st2}^{ot1}$ and $f_{st1}^{ot1}$ pairing channels implies the formation of Cooper pairs with similar spin. Given that the parent superconductor provides only opposite-spin (spin-singlet) amplitudes, the generation of same-spin correlations necessarily relies on spin-flip processes. In our model, the sd-exchange term serves as the sole mechanism for such spin-flipping. This is analytically confirmed by Eq. (\ref{eq:static_pairing_amp}), where both terms vanish in the limit $J_{sd} \rightarrow 0$. In this limit, the normal-state Hamiltonian reduces to an effective two-band odd-parity magnet, and the orbital subspace is spanned solely by the identity matrix. Consequently, the odd-frequency spin-triplet amplitude vanishes, leaving only the odd-parity spin-triplet $f_{st3}^{ot2}$ and the even-parity spin-singlet $f_{0,s}$. These results are consistent with previous theoretical predictions \cite{PhysRevB.111.144508}.

Up to this point, all nonvanishing pairing amplitudes belong to the orbital-triplet sector (even under orbital index exchange), as the Hamiltonian was restricted to intra-orbital couplings with degenerate bands. We further extend this analysis by introducing an inter-orbital hopping term \cite{minimal_model}:

\begin{equation}
    h_{pWM}^{'}({\mathbf{k}})= h_{pWM}({\mathbf{k}}) + 2 t_i \cos(k_x/2) \rho_x \otimes \sigma_0.
\label{eq:inter_orb_hopp}
\end{equation}

The introduction of inter-orbital hopping ($t_i$) facilitates the generation of orbital-singlet pairing correlations. These correlations are characterized by even-frequency, even-parity or odd-frequency, odd-parity along with spin-triplet symmetries, thereby expanding the landscape of superconducting pairings. Numerically, we find that for $J_{sd}, t_i \neq 0$, the non-zero pairing amplitudes can be classified into four distinct symmetry sectors as categorized in table \ref{tab:static_correlations}.

\begin{table}[h]
\caption{\label{tab:static_correlations}}
\begin{ruledtabular}
\footnotesize
\begin{tabular}{llp{5cm}}
Frequency & Parity & Allowed pairing channels \\ 
\colrule
Even & Even & $f_{ss}^{ot2}$, \quad $f_{ss}^{ot3}$, \quad $f_{st1}^{os}$ \\ 

Even & Odd & $f_{st2}^{ot1}$, \quad $f_{st3}^{ot2}$, \quad $f_{st3}^{ot3}$ \\

Odd & Even & $f_{st1}^{ot1}$ \\

Odd & Odd & $f_{st2}^{os}$ \\

\end{tabular}
\end{ruledtabular}
\end{table}

\section{Square Wave Periodic Driving}
\label{floquet}

Having established the properties of the static Hamiltonian, we next investigate the effects of periodic driving. Specifically, we introduce a square-wave modulation of the chemical potential, in which the system alternates between two values of it, $\mu_1$ for time $t_1$ and $\mu_2$ for time $t_2$, respectively, within each driving period. The total time period is $T=t_1+t_2$, such that $H(t+T)= H(t)$. The time-dependent Hamiltonian is defined as:
\begin{equation}
H(t) =
\begin{cases}
H_1= H_0- \mu_1\tau_z, & 0 \leq t < t_1, \\
H_2= H_0- \mu_2\tau_z, & t_1 \leq t < t_1+t_2=T,
\label{Ht}
\end{cases}
\end{equation}
where $H_0 = \mathcal{E}_{\mathbf{k}0} \tau_z\rho_0\sigma_0 + \alpha_{\mathbf{k}} \tau_0\rho_0\sigma_z + J_{sd} \tau_z\rho_z\sigma_x - d \tau_y\rho_0\sigma_y$ with $\mathcal{E}_{\mathbf{k}0}=-2t(\cos k_x+\cos k_y)$. The dynamics of such periodically driven systems are governed by Floquet theorem. The one-period time evolution operator (Floquet operator) is given as $U= \mathbb{T} e^{-i \int_0^T H(t)dt}$, where $\mathbb{T}$ denotes time-ordering operator. For the square wave driving, $U$ simplifies to a product of two independent unitary evolutions
\begin{equation}
U= e^{-i(H_0-\mu_2\tau_z)t_2} e^{-i(H_0-\mu_1\tau_z)t_1}.
\end{equation}

The stroboscopic dynamics is described by the effective Floquet Hamiltonian, $H_F = \frac{i}{T}\text{ln}(U)$. The eigenvalues of $H_F$ correspond to the quasi-energies $\epsilon_\alpha$, which are defined modulo the frequency of the Floquet Brillouin zone $2\pi/T$. While the asymmetric nature of the driving protocol potentially breaks the symmetries required to maintain a well-defined winding number, we utilize the gauge freedom of the starting time within the period $T$. By shifting to the symmetric time frames, we define two unitary equivalent Floquet operators, $U_1$ and $U_2$.

\begin{equation}
\begin{aligned}
    & U_1 = e^{-iH_1t_1/2} e^{-iH_2t_2} e^{-iH_1t_1/2}, \\
    & U_2 = e^{-iH_2t_2/2} e^{-iH_1t_1} e^{-iH_2t_2/2}.
\end{aligned}
\end{equation}

We then find the winding numbers $W_{1}$ and $W_{2}$ for these symmetric time frames. The FMZEM at the zero energy gap and the FMPEM appearing at the $\pi/T$ energy gap can be characterized by the following linear combinations.
\begin{equation}
    W_{0} = \frac{W_{1}+W_{2}}{2}, \qquad W_{\pi} = \frac{W_{1}-W_{2}}{2}.
\label{eq:floquet_winding_number}
\end{equation}

\begin{figure}
    \centering
    \includegraphics[width=1\linewidth]{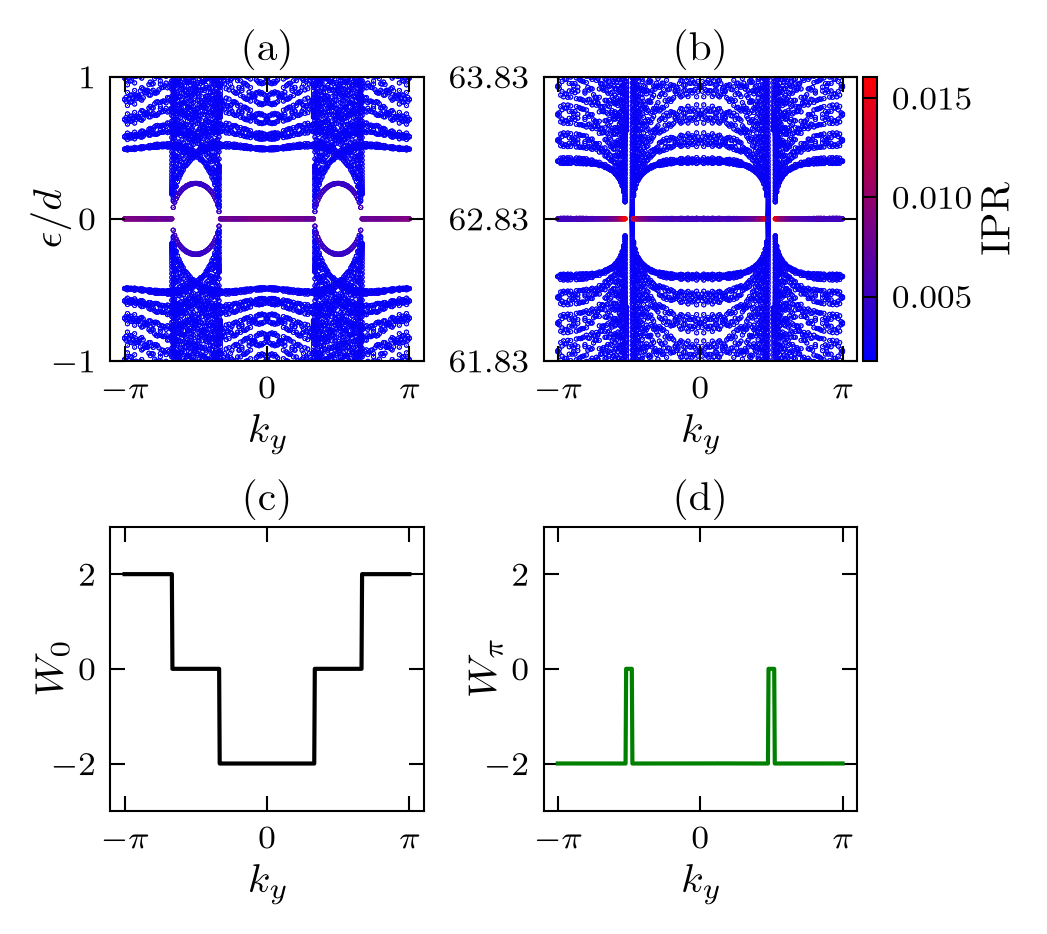}
    \caption{Depicts the Floquet quasi-energy ($\epsilon$) spectrum for the square-wave driving under a cylindrical geometry with open boundary conditions along the $x$-direction ($N_x = 600$) with periodic boundary conditions along the $y$-direction, and corresponding winding numbers. (a) Quasi-energy spectrum $\epsilon$ as a function of the transverse momentum $k_y$, highlighting the emergence of anomalous FMZEMs pinned at $\epsilon = 0$. (b) Shifting of the negative quasi-energy bands by $\Omega$ (where $\Omega = 2\pi/T$) to explicitly showcase the anomalous FMPEMs localized at the edge of the Floquet Brillouin zone ($\epsilon = \pm\pi/T$). (c) and (d) illustrate the transverse momentum-dependent Floquet winding numbers $W_0(k_y)$ and $W_{\pi}(k_y)$, respectively, characterizing the distinct topological protection of the zero- and $\pi$-energy boundary manifolds. The system parameters are fixed at $T = 1$, $t = 1$, $\mu = 3$, $d = 0.05$, $\alpha_x = 1$, $\alpha_y = 0$, and $J_{sd} = 1$.}
\label{fig:spectrum_sqwv}
\end{figure}

\begin{figure}
    \centering
    \includegraphics[width=1\linewidth]{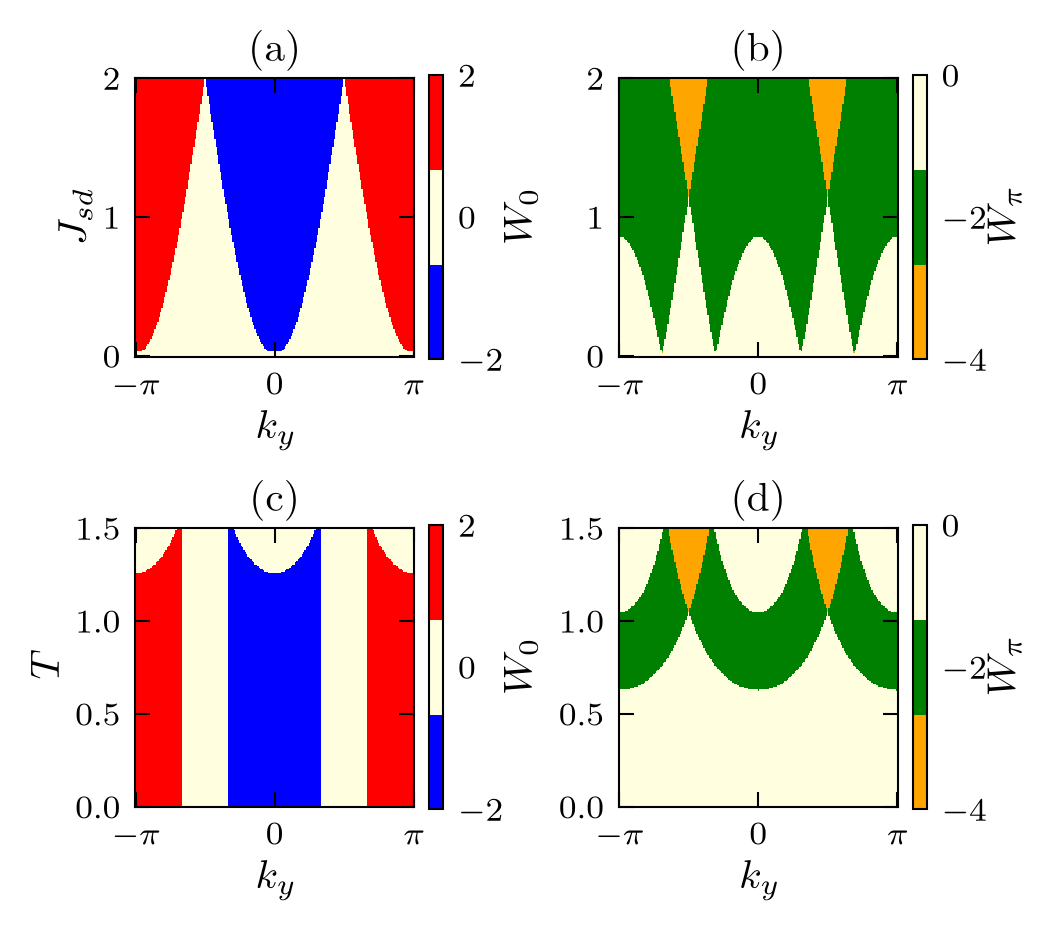}
    \caption{Showcases Floquet topological phase diagrams under the square-wave chemical potential driving protocol, illustrated via the sharp quantization of the dynamical invariants. (a) and (b) Winding numbers $W_0$ and $W_{\pi}$ mapped as a function of the sd coupling $J_{sd}$ and transverse momentum $k_y$, indexing the distinct topological phases hosting FMZEMs and FMPEMs, respectively; here, the driving period is fixed at $T = 1$. (c) and (d) Corresponding Floquet phase diagrams mapped as a function of the driving time period $T$, display the quantized values of $W_0$ and $W_{\pi}$ at a constant exchange coupling of $J_{sd} = 1$. The sharp boundaries between different color sectors explicitly denote phase transitions mediated by the closing and reopening of the bulk quasi-energy gaps. The remaining system parameters are the same as in Fig. \ref{fig:spectrum_sqwv}.}
\label{fig:ky_vs_jsd-tp_sqwv}
\end{figure}

For the remainder of this study, we set $t_1=t_2=T/2$ and $\mu_1 = -\mu_2=\mu$ throughout this paper unless otherwise specified. To visualize the emergent flat band FMZEM and FMPEM, we plot the quasi-energy spectrum of the effective Floquet Hamiltonian by imposing OBC along $x$ while maintaining $y$ periodic. The resulting spectrum (illustrated in Fig. \ref{fig:spectrum_sqwv} (a) and (b)) reveals the co-existence of conventional FMZEM at quasi-energy $\epsilon= 0$ and the anomalous FMPEM modes at quasi-energy $\epsilon= \hbar \Omega /2$. The topological nature of these edge modes is verified by the winding numbers $W_0$ and $W_\pi$, which within the bulk gaps take quantized values corresponding to quasienergies $0$ and $\pi$, respectively [Figs. \ref{fig:spectrum_sqwv} (c) and (d)]. In particular, $\lvert W_{0(\pi)}\rvert =  2$ signifies the existence of two pairs of Majorana modes at the corresponding $0$ and $\pi$-quasienergies. This observation is fully consistent with the bulk–boundary correspondence expected for a periodically driven system that belongs to the BDI symmetry class. In contrast to the static case, which is limited to two pairs of MZEMs, periodic driving facilitates the emergence of multiple pairs of $\pi$ Majorana modes in the low-frequency regime, see Fig. \ref{fig:ky_vs_jsd-tp_sqwv}. This is a direct consequence of the Floquet band folding within the quasi-energy Brillouin zone. Furthermore, in the high-frequency limit, a Floquet-Magnus expansion reveals that $H_F$ maps to the static Hamiltonian governed by the time-averaged chemical potential $(\mu_{avg} = 0)$. Consequently, FMZEMs appear analogous to the static MZEMs evaluated at $\mu = 0$. This correspondence (shown for $T=1$) is clearly observed by comparing the static spectrum in Fig.~\ref{fig:energy_and_winding_static}(a) with the driven spectrum in Fig.~\ref{fig:spectrum_sqwv}(a), highlighting the stroboscopic realization of the static topological phase.

Next, we examine the emergence of FMZEMs and FMPEMs as a function of exchange coupling $J_{sd}$ and the driving time period $T$. For analytical tractability, we restrict our focus to high symmetric points $(k_{x/y}= 0,\pm\pi)$. Since orbital degeneracy ensures that Majorana modes at each boundary always occur in pairs, analyzing a single orbital sector is sufficient to capture the topological phase transition points. Eliminating the orbital index $\rho$, the effective static BdG Hamiltonian at these high-symmetry points reduces to
\begin{equation}
    H_0 = \mathcal{E}_{\mathbf{k}0} \gamma_1 - d \gamma_2 + J_{sd} \gamma_3 ,
\end{equation}
where we define the composite matrices $\gamma_1= \tau_z\sigma_0$, $\gamma_2= \tau_y\sigma_y$ and $\gamma_3= \tau_z\sigma_x$. Notably, $\gamma_1$ and $\gamma_2$ anti-commute with each other while $\gamma_3$ commutes with both of them. Therefore, the term $J_{sd}$ can be considered as a scalar shift, which can be factored out of the Floquet operator as following
\begin{equation}
U = e^{- i J_{sd} \gamma_3 T} U_\phi,
\end{equation}
where $U_\phi$ is defined by:
\begin{equation}
    U_\phi = e^{-iH_{+} T/2} e^{-iH_{-} T/2}.
\end{equation}

Here $H_{\pm} = (\mathcal{E}_{\mathbf{k}0}\pm \mu) \gamma_1 - d \gamma_2$. Using the identity $e^{-i \boldsymbol{v_\pm}.\boldsymbol{\gamma}} = \cos(|\boldsymbol{v}_\pm|)\boldsymbol{I} - i\sin(|\boldsymbol{v}_\pm|)\hat{\boldsymbol{v}}_\pm.\boldsymbol{\gamma}$, and assuming that the eigenvalues of $U_\phi$ take the form $e^{-i\epsilon_\phi T}$, we derive the following transcendental equation 

\begin{equation}
\begin{aligned}
    \cos(\epsilon_\phi T)= & \cos\big(\frac{|\boldsymbol{v_+}|T}{2}\big) \cos\big(\frac{|\boldsymbol{v_-}|T}{2}\big) \\
    & - \sin\big(\frac{|\boldsymbol{v_+}|T}{2}\big) \sin\big(\frac{|\boldsymbol{v_-}|T}{2}\big) \frac{(\mathcal{E}_{\mathbf{k}0} ^2 - \mu^2 + d^2)}{|\boldsymbol{v_+}||\boldsymbol{v_-}|},
\end{aligned}
\label{eq:sqwv_zz1}
\end{equation}
where $|\boldsymbol{v_\pm}|= \sqrt{(\mathcal{E}_{\mathbf{k}0} \pm \mu)^2 + d^2}$. The quasi-energy expression is then given by:
\begin{equation}
    \epsilon = \pm J_{sd} + \epsilon_\phi \quad \text{mod} \quad\frac{2\pi}{T}.
\label{eq:sqwv_zz2}
\end{equation}
The phase boundaries for the FMZEM and FMPEM sectors are determined by the gap-closing condition at $\epsilon=0$  and $\epsilon= \pi/T$, respectively. These conditions translate to $\cos(\epsilon_\phi T)= \pm \cos(J_{sd} T)$ where the positive (negative) sign corresponds to the closing of the quasienergy gap at zero ($\pi/T$). In the high frequency limit $(T \rightarrow 0)$, Eq. (\ref{eq:sqwv_zz1}) reduces to the condition $|J_{sd}|^2= \mathcal{E}_{\mathbf{k}0}^2 + d^2$, exactly recovering the static topological phase boundaries derived in Section \ref{model}. To validate the analytical gap-closing conditions, we compute the winding numbers $W_{0}$ and $W_{\pi}$ across the $(k_y, J_{sd})$ parameter space, see Figs. \ref{fig:ky_vs_jsd-tp_sqwv} (a) and (b). By examining the high-symmetry lines at $k_y = 0$ and $k_y = \pm\pi$, we can directly compare the numerical transition points with our transcendental solutions. Substituting the parameters corresponding to Fig. \ref{fig:ky_vs_jsd-tp_sqwv} (a) and (b) into Eqs. (\ref{eq:sqwv_zz1}) and (\ref{eq:sqwv_zz2}), we obtain the critical sd couplings at which the quasienergy gap closes. For the zero-quasienergy sector, the gap-closing points occur at $|J_{sd}|_{[0]} \approx 0.033, 2.283, \ldots$, while for the $\pi$-quasienergy sector, the gap closes at $|J_{sd}|_{[\pi]} \approx 0.859, 3.108, \ldots$. Since the bulk quasienergy gap becomes extremely small for $J_{sd} > 2$, only the region with $J_{sd} < 2$ is displayed in the phase diagrams. Within this range, the analytically predicted transition points are in excellent agreement with the numerical emergence of FZMEMs and FPMEMs at the high-symmetry points. Furthermore, reflecting the intrinsic Floquet nature of the periodically driven system, the topological phase boundaries exhibit a characteristic periodicity of  $2\pi n$ $(n \in \mathcal{Z})$ in the $J_{sd}$ parameter.

We now examine the emergence of these Floquet Majorana modes as a function of the driving time period $T$. As illustrated in Figs. \ref{fig:ky_vs_jsd-tp_sqwv} (c) and (d), both FMZEMs and FMPEMs appear as the driving period increases. The phase transition boundaries can be determined analytically by identifying the parameters at which the Floquet quasienergy gap closes. At the high-symmetry points of the Brillouin zone, $\mathcal{E}_{\mathbf{k}0}$ takes only three distinct values, 0 and $\pm 4t$. Consequently, the quantities $(|\boldsymbol{v_+}|,|\boldsymbol{v_-}|)$ reduce to two inequivalent pairs, $(\Omega_0,\Omega_0)$ (for $X$ and $Y$) and  $(\Omega_\mp, \Omega_\pm)$ (for $\Gamma$ and $M$) where we define $\Omega_0 = \sqrt{\mu^2+d^2}$ and $\Omega_{\pm} = \sqrt{(4t \pm\mu)^2+d^2}$.
Substituting these expressions into Eq. (\ref{eq:sqwv_zz1}),  the transcendental equations defining the Floquet phase boundaries for the first pair reduce to:
\begin{equation}
     \cos(J_{sd}T)= \pm \Bigg[
    \cos^2 \big(\frac{T}{2}\Omega_0) - \sin^2 \big(\frac{T}{2}\Omega_0) \big(\frac{-\mu^2+d^2}{\mu^2+d^2}\big) \Bigg],
    \label{eq:first_pair}
\end{equation}
and for the second pair, it takes the form
\begin{align}
    \cos(&J_{sd}T)= \pm \Bigg[ \cos \big(\frac{T}{2}\Omega_+) \cos \big(\frac{T}{2}\Omega_-) \nonumber\\
    &- \sin \big(\frac{T}{2}\Omega_+) \sin \big(\frac{T}{2}\Omega_-) \big(\frac{(4t)^2 -\mu^2+d^2}{\Omega_+ \Omega_-}\big) \Bigg],
    \label{eq:second_pair}
\end{align}
where the +(-) sign on the right side in both equations corresponds to FMZEMs (FMPEMs).

Using the parameters of Fig. \ref{fig:ky_vs_jsd-tp_sqwv} (c) and (d), we solve Eqs. (\ref{eq:first_pair}) and Eq. (\ref{eq:second_pair}) by identifying the intersections between their left and right hand sides for driving periods up to $T=3.0$. We find that Eq. (\ref{eq:first_pair}) does not have solutions within this parameter range, indicating the absence of topological transitions associated with the points $X$ and $Y$. In contrast, Eq. (\ref{eq:second_pair}) yields a sequence of topological transitions originating from the $\Gamma$ and $M$ points, with zero quasi-energy gap closings at $T \approx 1.257, 2.094, 2.513$ and $\pi$-mode transitions at $T \approx 0.628, 1.047, 1.885$. In Figs. \ref{fig:ky_vs_jsd-tp_sqwv} (c) and (d), we consider the driving period to $T \le 1.5$, as the quasienergy gap becomes exceedingly small for larger values of $T$. Within this range, the analytically predicted transition points along high symmetry points are in exact agreement with the quasienergy gap closings obtained from direct numerical diagonalization of the Floquet operator. An analogous analysis has been performed for the delta-kick driving protocol. The corresponding derivation, phase-boundary equations, and symmetry considerations are presented in Appendix~\ref{delta_kick}.

\section{Transport Properties of FMZEM and FMPEM}
\label{floquet transport}

In the non-equilibrium regime induced by periodic driving, the continuous time translation symmetry is broken down to a discrete symmetry $H(t+T) = H(t)$. This allows for a Fourier decomposition of the time-dependent Hamiltonian into discrete frequency components:

\begin{equation}
H_{(m)} = \frac{1}{T} \int_0^T H(t) e^{im\Omega t} dt,
\end{equation}

where $\Omega = 2\pi/T$ is the driving frequency. This transformation maps the time-dependent problem onto a static lattice in frequency space, known as the Sambe space representation \cite{Sambe}. In this extended Hilbert space (explicitly discussed in Appendix \ref{extended_space}), the elements of the Floquet Hamiltonian are defined as:

\begin{equation}
    [\mathbf{H}^{\text{ext}}_F]_{mn} = \mathbf{H}_{(m-n)} + m\Omega \delta_{mn} \mathbf{I},
\end{equation}
where $m, n$ are Floquet mode indices representing the number of photons exchanged with the driving field. The diagonal terms $H_{(0)} + m\Omega$ correspond to the onsite energies of the $m$-th photon sector, while the off-diagonal terms $H_{(m-n)}$ describe the hopping between sectors via the absorption or emission of $\lvert m-n\rvert$ photons.

For numerical implementation, the infinite-dimensional Sambe space is truncated to a finite number of photon sectors, $m \in[-M,M]$. Following a similar system-lead configuration as in the static case, we define the retarded Floquet Green's function in this extended space:

\begin{equation}
    \mathbf{G}_F ^R(E)= [(E + i \eta)\mathbf{I}_F - \mathbf{H}_F - \mathbf{\Sigma}_F]^{-1},
\end{equation} 
where $\mathbf{I}_F$ is the identity matrix of the truncated Sambe space. Within the WBL, the lead self-energy $\mathbf{\Sigma}$ is assumed to be frequency-independent. Consequently, the self-energy matrix in the extended Floquet space simplifies to a block-diagonal form:
\begin{equation}
    \mathbf{\Sigma}_F= \mathbf{\Sigma} \otimes\mathbf{I}_{[2M+1]}.
\end{equation}

This construction allows us to evaluate the time-averaged transport properties by summing over the contributions of all relevant Floquet sidebands. The transport in periodically driven systems fundamentally differs from that in static systems due to inelastic scattering processes. An electron incident from the lead at energy $E$ can be scattered into energies $E \pm n\Omega$, corresponding to the $n$-th Floquet sideband, via the absorption or emission of $n$ photons. To accurately describe the conductance, one must account for all scattering channels connecting the incident $0$-th photon sector to the outgoing $n$-th sectors.

\begin{figure}
    \centering
    \includegraphics[width=1\linewidth]{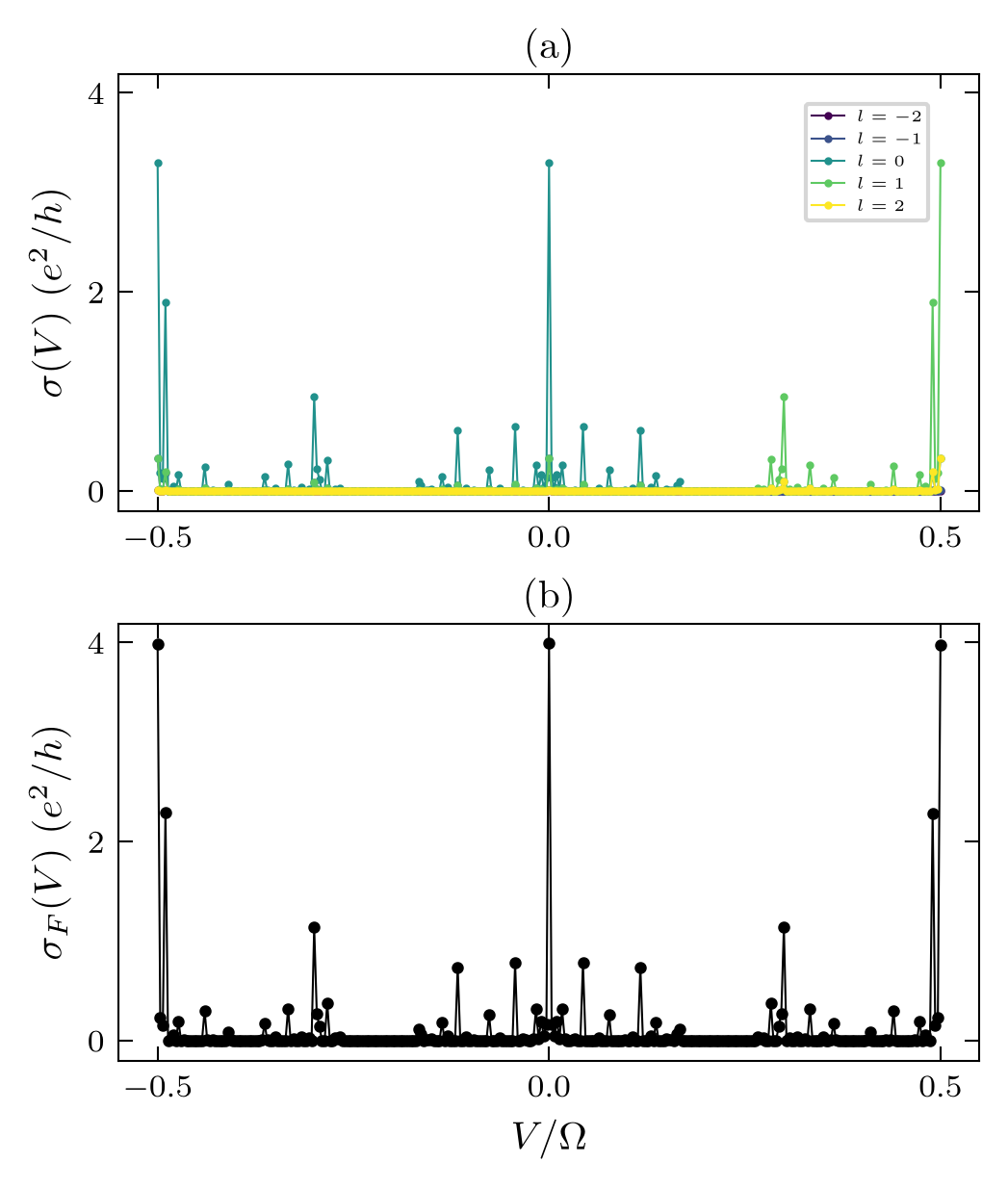}
    \caption{Differential conductance in square-wave periodic driving as a function of bias voltage $(V)$ with cylindrical geometry ($N_x = 800$) for $k_y = 0$. In (a), we have plotted the $\sigma(V)$ contributions from different individual photon sectors from $l = - 2$ to $l = +2$, and in (b), the total summed conductance $\sigma_F (V)$, which clearly shows quantized peaks of $4e^2/h$ corresponding to $V= 0$ and $V= \Omega/2$. The remaining system parameters are the same as in Fig. \ref{fig:spectrum_sqwv} with the interface barrier strength of $\Gamma_0 = 0.2$.}
\label{fig:transport_floquet}
\end{figure}

The non-zero local reflection block (similar to Eq. (\ref{rc})) is derived from the $(n,0)$-th block of the retarded Floquet Green's function, which maps the $0$-th sector to the $n$-th sector as follows:
\begin{equation}
    \textbf{r}_c ^{(n,0)} (E) = \delta_{n,0} \mathbf{I}_c - i \Gamma_0 \mathbf{G}_c^{R,(n,0)},
\end{equation}
where the subscript $c$ represents the contact block, the $\delta_{n,0}$ reflects the condition that the incident electron enters exclusively through the $0$-th sector. The corresponding normal and local Andreev reflection probabilities are $R_{ee}^{(n)} = \text{Tr}[\textbf{r}_{ee}^{(n,0) \dagger} \textbf{r}_{ee}^{(n,0)}]$ and $R_{he}^{(n)} = \text{Tr}[\textbf{r}_{he}^{(n,0) \dagger}\textbf{r}_{he}^{(n,0)}]$ respectively. Summing over all outgoing channels, the time-averaged differential conductance at bias $V$ follows a modified BTK formula:
\begin{equation}
    \sigma(V) = \frac{e^2}{h} \sum_{n} \left[ \delta_{n,0} N - R_{ee}^{(n)}(V) + R_{he}^{(n)}(V) \right].
\end{equation}
where $N$ is the number of transverse channels. As noted in previous studies, the bare conductance $\sigma (V)$ often fails to exhibit the expected topological quantization due to the redistribution of spectral weight across Floquet replicas. To recover the quantized signature, we employ the Floquet sum rule \cite{floquet_sum_rule}, which accounts for the total contribution of all quasi-energy states shifted by integer multiples of the drive frequency $\Omega$:

\begin{equation}
    \sigma_{F}(V) = \sum_l \sigma(V + l\Omega).
\end{equation}
We illustrate this approach for a system in cylindrical geometry. Fig. \ref{fig:transport_floquet} shows the differential conductance as a function of the bias voltage $V$ for a representative transverse momentum $k_y = 0$. Although the conductance contributions from the individual Floquet photon sectors are not quantized, their sum, corresponding to the total Floquet conductance $\sigma_{F}$, exhibits clear quantized signatures of the underlying topological phase. In particular, the quantized conductance peak at $V = 0$ signifies the presence of two pairs of FMZEMs, while the quantized peaks at $V= \Omega/2$ provide evidence for two pairs of FMPEMs. The restoration of conductance quantization through the Floquet sum rule therefore constitutes a robust transport signature of Floquet Majorana modes in the system.

\section{Cooper Pair Correlations in the Non-Equilibrium Regime}
\label{floquet pairing}
In periodically driven systems, the anomalous Green's function is naturally defined in the extended Floquet (Sambe) space. The matrix element $F^{n,m}(\mathbf{k},z)$ describes the pairing of two electrons accompanied by the exchange of $|n-m|$ photons. As a result, the standard fermionic antisymmetry condition written in Eq. (\ref{F}) must be extended to explicitly account for the Floquet (Sambe) indices. Following the decomposition of the time-dependent anomalous Green's function \cite{RevModPhys.86.779}, we write

\begin{equation}
\begin{aligned}
    & F_{\rho_1,\rho_2;\sigma_1,\sigma_2}(\mathbf{k}_1,\mathbf{k}_2;t_1,t_2) \\
    & = \int \frac{dz}{2\pi}\sum_{n,m}  e^{-i(z+n\Omega)t_1} e^{+i(z+m\Omega)t_2} F^{n,m}_{\rho_1,\rho_2;\sigma_1,\sigma_2}(\mathbf{k}_1,\mathbf{k}_2;z),
\end{aligned}
\end{equation}
where $z= \omega+i\eta$, $\eta \rightarrow 0$. Under the simultaneous exchange of all degrees of freedom—orbital ($\rho$), spin ($\sigma$), momentum ($\mathbf{k}$) and time ($t$), the transformed function reads:

\begin{equation}
\begin{aligned}
    & F_{\rho_2,\rho_1;\sigma_2,\sigma_1}(\mathbf{k}_2,\mathbf{k}_1;t_2,t_1) \\
    & = \int \frac{dz}{2\pi}\sum_{n,m}  e^{-i(z+n\Omega)t_2} e^{+i(z+m\Omega)t_1} F^{n,m}_{\rho_2,\rho_1;\sigma_2,\sigma_1}(\mathbf{k}_2,\mathbf{k}_1;z) \\
    & = \int \frac{dz}{2\pi}\sum_{n,m}  e^{-i(z+n\Omega)t_1} e^{+i(z+m\Omega)t_2} F^{-m,-n}_{\rho_2,\rho_1;\sigma_2,\sigma_1}(\mathbf{k}_2,\mathbf{k}_1;-z),  \\
\end{aligned}
\end{equation}
where we have utilized the substitutions $z \rightarrow -z$, $n \rightarrow -n$, and $m \rightarrow -m$ and then relabeled the dummy indices $n \leftrightarrow m$ to realign the exponential kernels. Comparing these expressions, we conclude that periodic driving introduces a specific antisymmetry constraint on the Floquet indices: the exchange of indices must follow the mapping $(n,m) \rightarrow (-m, -n)$. Consequently, the Cooper pair amplitude must satisfy:
\begin{equation}
F^{n,m}_{\rho_1,\rho_2;\sigma_1,\sigma_2}(\mathbf{k}_1,\mathbf{k}_2;z) = - F^{-m,-n}_{\rho_2,\rho_1;\sigma_2,\sigma_1}(\mathbf{k}_2,\mathbf{k}_1;-z),
\end{equation}

where the inversion of the Floquet indices reflects the exchange of the two time arguments under fermionic antisymmetry. The coupling between different Floquet sectors can then redistribute the symmetry among frequency, momentum, spin, orbital, and Floquet-index channels, allowing odd-frequency pairing components to emerge even when the equilibrium pairing is purely even-frequency.
This symmetry classification is crucial for understanding the nature of superconducting order in the presence of periodic fields.

\subsection*{Symmetry classification of Floquet Cooper pairs}
The application of a periodic drive introduces a manifold of Floquet sidebands, which substantially expands the space of superconducting pair correlations.  To characterize these emergent symmetries, we numerically evaluate the Floquet anomalous Green's function for time-dependent Hamiltonian (see Eq. (\ref{Ht})) with non-zero inter-orbital hopping written in Eq.(\ref{eq:inter_orb_hopp}). We find that periodic driving doubles the number of symmetry-distinct Cooper-pair amplitudes relative to the equilibrium case.

\subsubsection{Even-Floquet Correlations}
The first class of correlations corresponds to the even-Floquet pairing ($F^{n,m}=F^{-m,-n}$). In this sector, the pairing symmetries mirror those of the static case, where the total wave function satisfies the standard Berezinskii constraint across frequency, spin, orbital, and parity (momentum) degrees of freedom. These are categorized in Table \ref{tab:even_flq_correlations}.

\begin{table}[h]
\caption{\label{tab:even_flq_correlations}}
\begin{ruledtabular}
\footnotesize
\begin{tabular}{lllp{4cm}}
Floquet & Frequency & Parity & Allowed pairing channels \\ 
\colrule
Even & Even & Even & $f_{ss}^{ot2}$, \quad $f_{ss}^{ot3}$, \quad $f_{st1}^{os}$ \\ 

Even & Even & Odd & $f_{st2}^{ot1}$, \quad $f_{st3}^{ot2}$, \quad $f_{st3}^{ot3}$ \\

Even & Odd & Even & $f_{st1}^{ot1}$ \\

Even & Odd & Odd & $f_{st2}^{os}$ \\

\end{tabular}
\end{ruledtabular}
\end{table}

\subsubsection{Odd-Floquet Correlations and Frequency Inversion}
The second class involves an odd-Floquet pairing ($F^{n,m}=-F^{-m,-n}$). Here, we observe a systematic inversion of the frequency index, while the parities of the spatial momentum, orbital, and spin degrees of freedom remain preserved relative to the static limit. The resulting correlations are categorized in Table \ref{tab:odd_flq_correlations}.

\begin{table}[h]
\caption{\label{tab:odd_flq_correlations}}
\begin{ruledtabular}
\footnotesize
\begin{tabular}{lllp{4cm}}
Floquet & Frequency & Parity & Allowed pairing channels \\ 
\colrule
Odd & Odd & Even & $f_{ss}^{ot2}$, \quad $f_{ss}^{ot3}$, \quad $f_{st1}^{os}$ \\ 

Odd & Odd & Odd & $f_{st2}^{ot1}$, \quad $f_{st3}^{ot2}$, \quad $f_{st3}^{ot3}$ \\

Odd & Even & Even & $f_{st1}^{ot1}$ \\

Odd & Even & Odd & $f_{st2}^{os}$ \\

\end{tabular}
\end{ruledtabular}
\end{table}

The physical origin of this doubling lies in the nature of the drive. Since our Floquet protocol consists of a periodic modulation of the chemical potential, it inherently respects spatial inversion, spin-rotation, and orbital-basis symmetries. Consequently, to satisfy the fundamental fermionic antisymmetry constraint, any exchange in the "Odd-Floquet" sector must be compensated by a flip in the frequency parity. This result highlights the potential of periodic driving to act as a symmetry converter; specifically, it allows for the dynamic generation of odd-frequency states from conventional even-frequency precursors.

\section{Robustness against onsite disorder}
\label{disorder}
\begin{figure}
    \centering
    \includegraphics[width=1\linewidth]{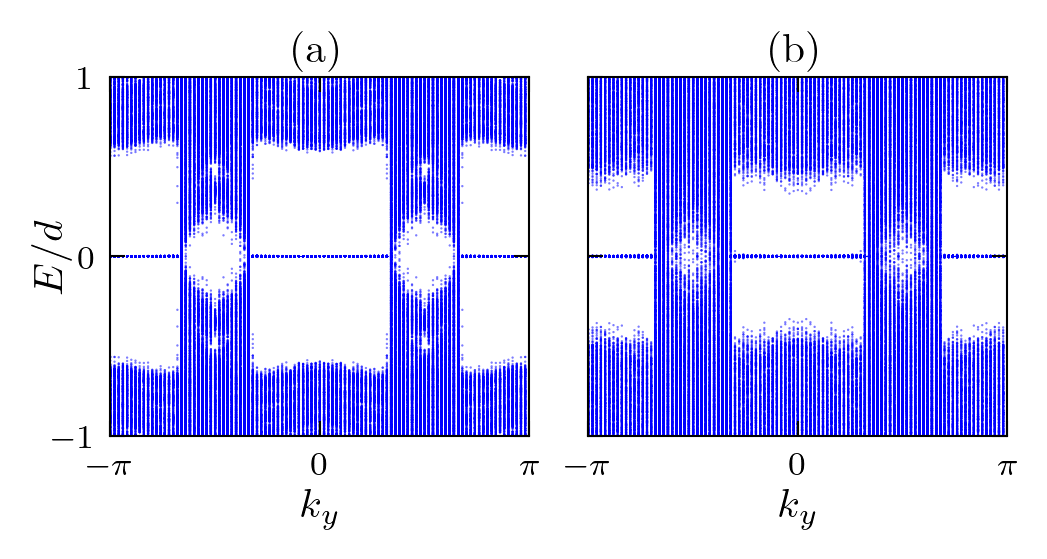}
    \caption{Disordered energy spectrum by superimposing $100$ independent random disorder realizations on cylinder geometry with $N_x=400$ as a function of transverse momentum $k_y$, (a) disorder strength $\mu_{\text{dis}}^{\text{max}} = 0.25$ and (b) disorder strength $\mu_{\text{dis}}^{\text{max}} = 0.5$. The remaining system parameters are the same as in Fig. \ref{fig:energy_and_winding_static}.}
\label{fig:static_disorder}
\end{figure}

To assess the experimental feasibility of our proposal, we investigate the robustness of the Majorana edge modes against spatial disorder. We introduce a static, random on-site disorder into the chemical potential of the Hamiltonian as follows
\begin{equation}
    [V_{\text{dis}}]_{ij} = \delta_{ij} \mu_{\text{dis}}^i (\tau_z \otimes \rho_0 \otimes \sigma_0),
\label{eq:disorder_form}
\end{equation}
where $\mu_{\text{dis}}^i$ represents the local disorder strength at the $i$-th lattice site, , sampled from a uniform distribution $[-\mu_{\text{dis}}^{\text{max}}/2,\mu_{\text{dis}}^{\text{max}}/2]$. The inclusion of the Nambu matrix $\tau_z$ ensures that the fundamental PHS constraint is satisfied, allowing the electron and hole sectors to experience disorder potentials of opposite signs ($\mu_{\text{dis}}$ and $-\mu_{\text{dis}}$, respectively). To maintain the efficiency of our momentum-space analysis, we implement strip disorder. Random chemical potentials are added along the $x$-direction while maintaining spatial uniformity along the $y$-direction. This configuration preserves translational invariance along the $y$-axis, ensuring that the transverse momentum $k_y$ remains a valid quantum number. The disordered Hamiltonian is given by
\begin{equation}
    H_{\text{dis}}(x,k_y)= H_{\text{clean}}(x,k_y) +  \sum_{i=1}^{N_x}   \Psi_{i,k_y}^{\dagger} [\mu_{\text{dis}}^i (\tau_z \otimes \rho_0 \otimes \sigma_0)] \Psi_{i,k_y}
\end{equation}
By monitoring the energy spectrum as a function of the disorder amplitude $\mu_{\text{dis}}^{\text{max}}$, we analyze the topological protection of the MZEMs. These states remain strictly pinned to zero energy even in the presence of a significantly strong disorder. Fig. \ref{fig:static_disorder} presents the spectrum for strong disorder regimes ($\mu_{\text{dis}}^{\text{max}}$ = 0.25) and ($\mu_{\text{dis}}^{\text{max}}$ = 0.5), respectively, clearly illustrating the persistence of the zero-energy manifold.
The stability of these modes against this specific form of disorder($V_{\text{dis}} \propto \tau_z\rho_0\sigma_0$) comes from the fact that the disorder operator anti-commutes with both the CS operator $\Gamma$ and the PHS operator $\mathcal{C}$. This anti-commutation prevents the mixing of edge states with different chiralities, thereby prohibiting any shift in quasi-energy and ensuring the robust quantization of the Majorana states within the topological regime.

\begin{figure}
    \centering
    \includegraphics[width=1\linewidth]{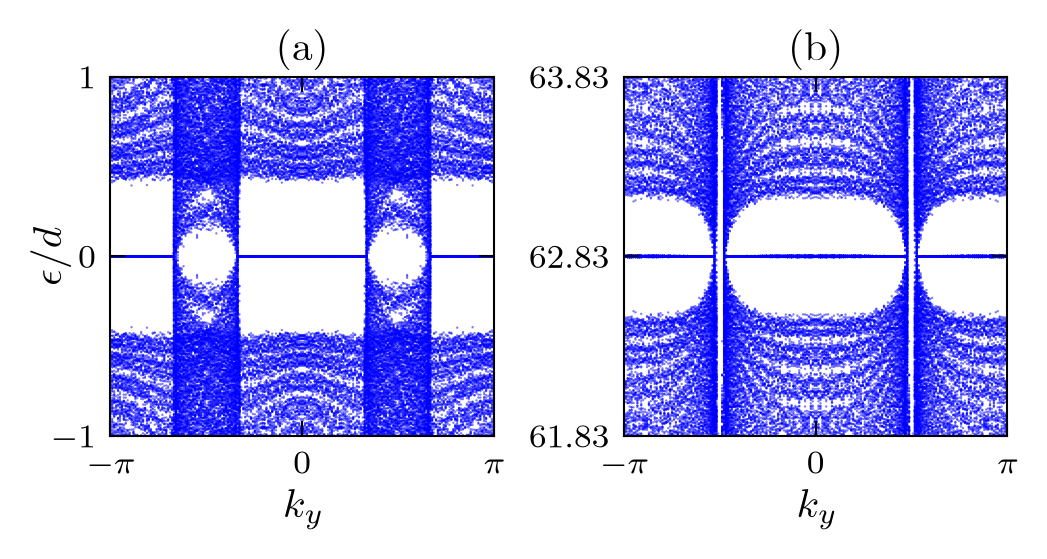}
    \caption{Disordered quasi-energy spectrum by superimposing $10$ independent random disorder realizations on cylinder geometry with $N_x=600$ as a function of transverse momentum $k_y$ with strong disorder strength $\mu_{\text{dis}}^{\text{max}} = 0.25$, (a) quasi-energy spectrum near $\epsilon=0$ and (b) shifted quasi-energy spectrum near $\epsilon=\Omega/2$ (negative quasi-energies shifted by $\Omega$ showcasing $\pi$ modes). The remaining system parameters are the same as in Fig. \ref{fig:spectrum_sqwv}.}
\label{fig:floquet_disorder}
\end{figure}

In the non-equilibrium periodically driven regime, we further investigate the robustness of the Floquet topological phases by introducing static on-site disorder independently into each driving step within a single period $T$. Remarkably, both FMZEMs and FMPEMs remain robust against strong disorder, persisting until the protecting bulk quasienergy gaps eventually close. This robustness is illustrated in Fig.~\ref{fig:floquet_disorder}, which shows the quasienergy spectrum for a representative disorder strength of $(\mu_{\text{dis}}^{\text{max}} = 0.25)$. Although bulk states undergo noticeable broadening and spectral fluctuations, the zero- and $\pi$-quasienergy edge modes remain sharply localized within their respective gaps. The robustness of these edge states originates from the symmetry-protected nature of the Floquet topological phase. In particular, the FMPEMs, residing at the boundary of the Floquet Brillouin zone, are protected by the same particle-hole and chiral symmetries that stabilize the zero-energy Majorana modes. Since the applied driving protocol preserves these symmetries even in the presence of random on-site potential fluctuations, neither the zero-modes nor the $\pi$-modes can hybridize with the bulk spectrum as long as the corresponding quasienergy gaps remain open. The persistence of these topological edge modes in the presence of substantial spatial disorder highlights the experimental robustness of the proposed Floquet topological phase and its potential realization in realistic condensed-matter systems, where disorder is unavoidable.

\section{Conclusion}
\label{conclusion}
In this work, we have theoretically investigated the equilibrium and non-equilibrium topological properties of a two-dimensional heterostructure comprising a minimal $p$-wave magnetic system proximitized by an $s$-wave superconductor. In the static limit, the BdG Hamiltonian belongs to the BDI symmetry class and hosts MZEMs when confined along the $x$-direction. We identified seven distinct nodal topological phases, each characterized by bulk gap closings and quantized chiral winding numbers. The existence of these Majorana modes was established through their real-space localization and further corroborated by the quantized zero-bias conductance obtained within the BTK formalism. By transforming the Hamiltonian into the band basis, we demonstrated that the coexistence of $p$-wave magnetism and conventional $s$-wave superconductivity naturally generates an effective $p$-wave pairing channel. A comprehensive symmetry analysis of the anomalous Green's function revealed a remarkably rich superconducting landscape. In addition to intrinsic even-frequency spin-singlet pairing, the system supports both odd-parity even-frequency and even-parity odd-frequency spin-triplet correlations. Furthermore, finite inter-orbital hybridization gives rise to an unconventional odd-frequency, odd-parity, orbital-singlet pairing channel, highlighting the intricate interplay between orbital, spin, and frequency degrees of freedom.

We subsequently generalized our analysis to periodically driven systems using square-wave and Delta-kick modulations of the chemical potential. Floquet engineering enables a controlled transition from a static phase to dynamically generated topological phases supporting both Floquet zero-energy and anomalous $\pi/T$-energy flat bands. These phases were characterized by winding numbers defined in symmetric time frames, while their experimentally observable transport signatures were shown to obey the Floquet sum rule, providing a direct route for their detection. One of the central findings of this work is that periodic driving fundamentally enriches both the topological and superconducting properties of the heterostructure. The Floquet drive generates multiple nodal points, enabling higher winding numbers and a corresponding proliferation of Majorana flat bands. Simultaneously, the additional Floquet degree of freedom doubles the number of symmetry-allowed Cooper-pair correlations within the generalized Berezinskii classification. Although even Floquet sectors retain the equilibrium pairing symmetries, odd Floquet sectors host complementary frequency-converted pairing channels, thereby substantially expanding the landscape of unconventional superconductivity accessible in driven systems. Finally, we demonstrate that both equilibrium and Floquet Majorana edge modes remain robust against strong spatial disorder, surviving until the protecting bulk and quasienergy gaps close. The persistence of both zero- and $\pi$-Majorana modes under realistic disorder conditions underscores the experimental feasibility of the proposed platform.

Overall, our work establishes periodically driven $p$-wave magnetic/$s$-wave superconducting heterostructures as a versatile platform in which topology, magnetism, superconductivity, and Floquet dynamics cooperate to generate robust Majorana excitations and a diverse spectrum of unconventional Cooper-pair symmetries. Beyond providing experimentally accessible signatures through quantized transport, our results reveal Floquet engineering as a powerful route for tailoring topological superconductivity and designing nonequilibrium quantum phases with enhanced symmetry structures. These findings open promising directions toward the controlled realization of Majorana-based quantum devices and the exploration of engineered unconventional superconductivity in  quantum materials.

\appendix
\section{Delta-Kick Driving}
\label{delta_kick}

We now consider a periodic driving protocol where the chemical potential is modulated via a sequence of Dirac delta kicks. The time-dependent Hamiltonian is given by:
\begin{equation}
H(t)= H_0 - \mu \sum_n \delta(t-nT) \tau_z,
\end{equation}
where $T$ denotes the time period of the drive and $\mu$ represents the kick strength. The corresponding Floquet operator, which describes the evolution over a single period, decomposes into a free evolution term and a phase rotation:
\begin{equation}
U= e^{i \mu\tau_z} e^{-iH_0T}.
\end{equation}

The structure of $U$ reveals a fundamental $2\pi$ periodicity with respect to the kick strength $\mu$. Notably, for $\mu = n\pi$, the operator $e^{i \mu\tau_z}$ reduces to $(-1)^n \mathbf{I}$. In this limit, the drive contributes only a global phase, leaving the physical eigenstates and the quasi-energy spectrum identical to the static case (up to a possible shift of $\pi/T$ in the quasi-energy). A shift of $\mu \rightarrow \mu + \pi$ results in $U \rightarrow -U$, effectively inverting the Floquet spectrum. This transformation maps FMZEMs at $\mu$ to FMPEMs at $\mu + \pi$, and vice-versa. Furthermore, at $\mu = \pi/2$, the identity component of the kick operator vanishes ($\cos{\mu}=0$), maximizing the non-trivial unitary evolution. Crucially, this periodicity in $\mu$ is an intrinsic property of the kick operator and remains independent of the driving period $T$.

As with the square-wave protocol, the Floquet Hamiltonian $H_F = \frac{i}{T}\ln{U}$ derived from the standard time frame generally breaks the symmetries required to define a robust topological invariant (such as the winding number). To restore these symmetries, we perform a gauge transformation to symmetric time frames. We define two unitarily equivalent operators, $U_1$ and $U_2$, given by:

\begin{align}
    U_1 &= e^{-iH_0T/2} e^{i\mu\tau_z} e^{-iH_0T/2}, \\
    U_2 &= e^{i\mu\tau_z/2} e^{-iH_0T} e^{i\mu\tau_z/2}.
\end{align}
We follow the same procedure as in the section \ref{floquet} to calculate winding numbers $W_{0}$ and $W_{\pi}$, which characterize the FMZEMs and FMPEMs. 

\begin{figure}
    \centering
    \includegraphics[width=1\linewidth]{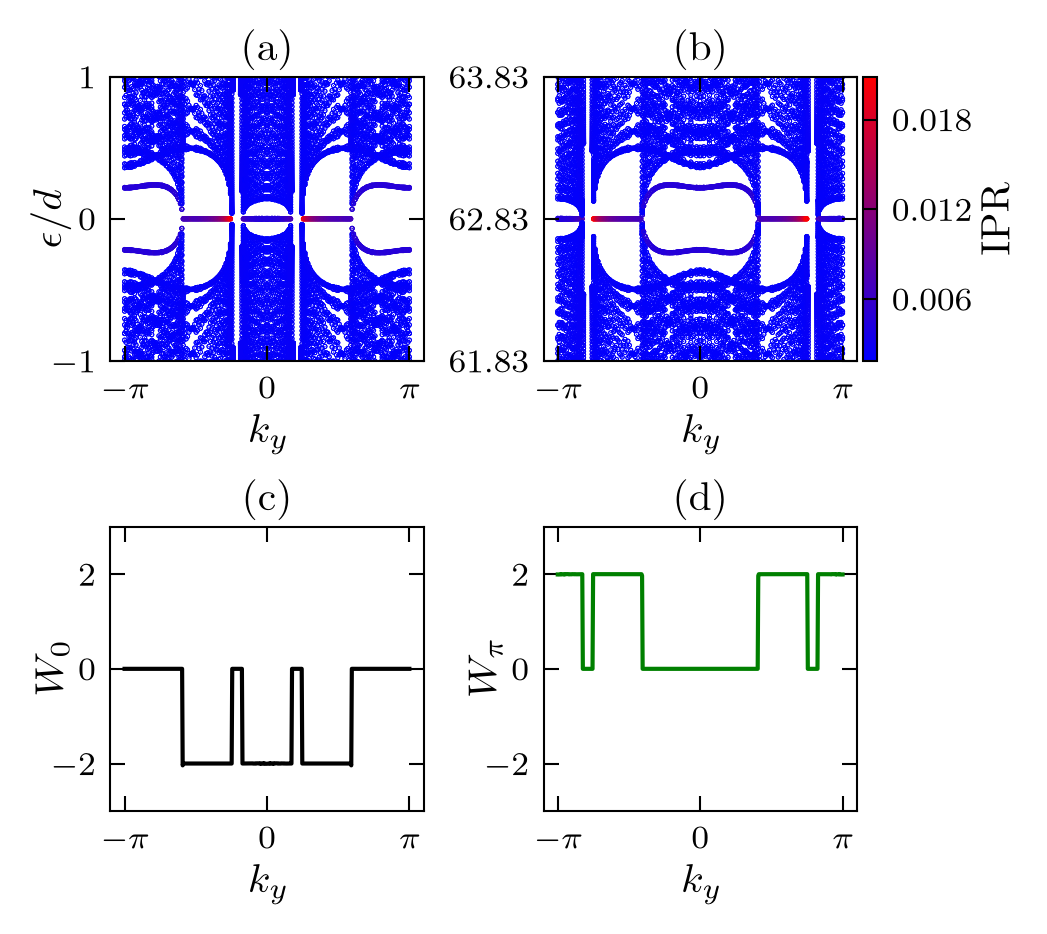}
    \caption{Depicts the Floquet quasi-energy ($\epsilon$) spectrum for the delta-kick driving under a cylindrical geometry with open boundary conditions along the $x$-direction ($N_x = 600$) with periodic boundary conditions along the $y$-direction and corresponding winding numbers. (a) Quasi-energy spectrum $\epsilon$ as a function of the transverse momentum $k_y$, highlighting the emergence of anomalous FMZEMs pinned at $\epsilon = 0$. (b) Shifting of the negative quasi-energy bands by $\Omega$ (where $\Omega = 2\pi/T$) to explicitly showcase the anomalous FMPEMs localized at the edge of the Floquet Brillouin zone ($\epsilon = \pm\pi/T$). (c) and (d) illustrate the transverse momentum-dependent Floquet winding numbers $W_0(k_y)$ and $W_{\pi}(k_y)$, respectively, characterizing the distinct topological protection of the zero- and $\pi$-energy boundary manifolds. The system parameters are fixed at $T = 1$, $t = 1$, $\mu = \pi/2$, $d = 0.05$, $\alpha_x = 1$, $\alpha_y = 0$, and $J_{sd} = 1$.}
\label{fig:delk_spectrum}
\end{figure}

To analytically track the topological phase transitions, we focus on the high-symmetry points in the Brillouin zone $(k_{x/y} = 0, \pm \pi)$, where the term $\alpha_{\mathbf{k}}$ vanishes. In this subspace, the static Hamiltonian reduces to:

\begin{equation}
    H_0 = \mathcal{E}_{\mathbf{k}0} \gamma_1 - d \gamma_2 + J_{sd} \gamma_3.
\end{equation}

Following the algebraic approach utilized for square-wave driving, the Floquet operator can be decomposed as $U= e^{- iJ_{sd} \gamma_3 T} U_\phi$ where the operator $U_\phi$ encapsulates the combined effect of the delta kick and the remaining static terms:

\begin{equation}
    U_\phi = e^{i\mu \gamma_1} e^{-i[\mathcal{E}_{\mathbf{k}0} \gamma_1 - d \gamma_2] T}. 
\end{equation}

By employing the general rotation identity for Pauli-like matrices used in the previous section, we derive the characteristic equation for the quasi-eigenvalues $\epsilon_\phi$ associated with $U_\phi$:
\begin{equation}
\begin{aligned}
    \cos(\epsilon_\phi T)=  & \cos\big(\mu \big) \cos\big(T\sqrt{\mathcal{E}_{\mathbf{k}0}^2 + d^2}\big) \\
     & + \frac{\mathcal{E}_{\mathbf{k}0}}{\sqrt{\mathcal{E}_{\mathbf{k}0}^2 + d^2}} \sin\big(\mu \big) \sin\big(T\sqrt{\mathcal{E}_{\mathbf{k}0}^2 + d^2}\big).
\end{aligned}
\end{equation}

The total quasi-energy of the system is given by the relation $\epsilon = \pm J_{sd} + \epsilon_\phi$. The phase transition points where the quasi-energy gap closes at either $\epsilon = 0$ or $\epsilon = \pi/T$ are then precisely determined by the condition:

\begin{equation}
\begin{aligned}
    \cos(J_{sd} T)= \pm \Bigg[ & \cos\big(\mu \big) \cos\big(T\sqrt{\mathcal{E}_{\mathbf{k}0}^2 + d^2}\big) \\
     & + \frac{\mathcal{E}_{\mathbf{k}0}}{\sqrt{\mathcal{E}_{\mathbf{k}0}^2 + d^2}} \sin\big(\mu \big) \sin\big(T\sqrt{\mathcal{E}_{\mathbf{k}0}^2 + d^2}\big)\Bigg].
\end{aligned}
\end{equation}

\begin{figure}
    \centering
    \includegraphics[width=1\linewidth]{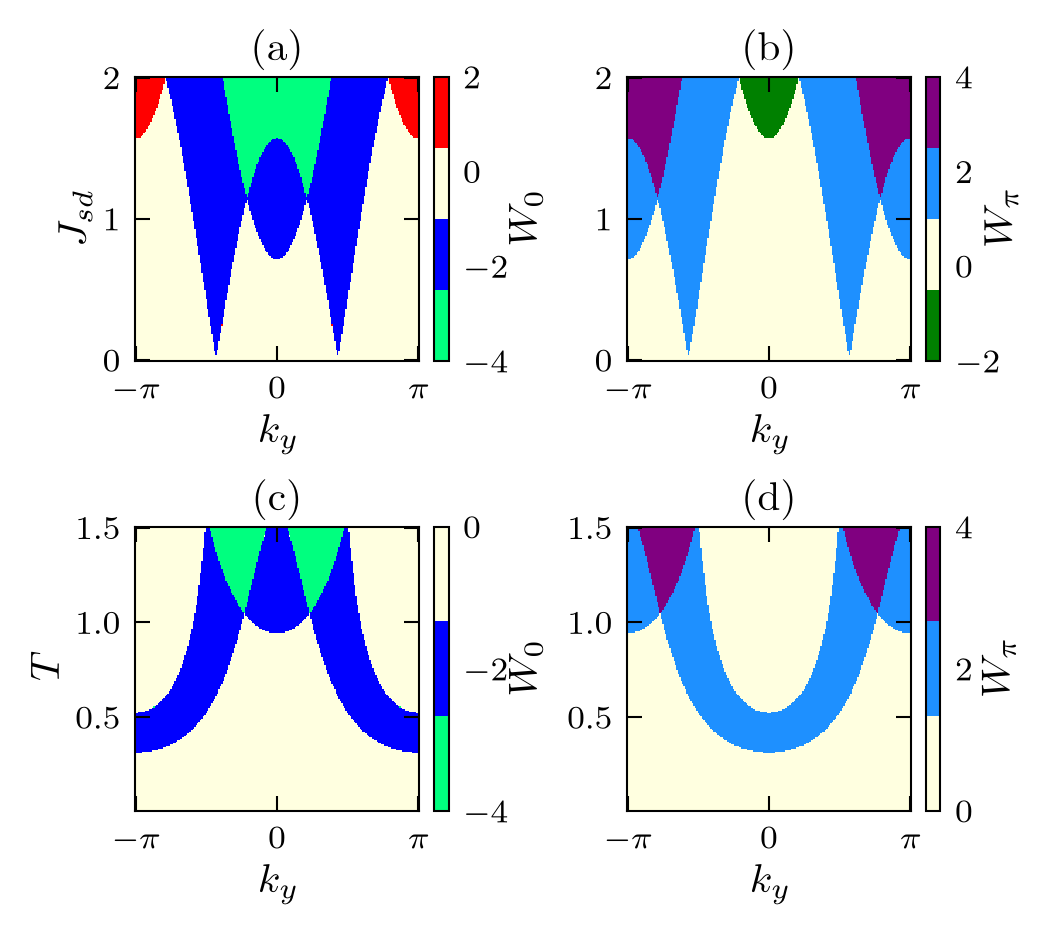}
    \caption{Showcases Floquet topological phase diagrams under the delta-kick chemical potential driving protocol, illustrated via the sharp quantization of the dynamical invariants. (a) and (b) Winding numbers $W_0$ and $W_{\pi}$ mapped as a function of the sd coupling $J_{sd}$ and transverse momentum $k_y$, indexing the distinct topological phases hosting anomalous FMZEMs and FMPEMs, respectively; here, the driving period is fixed at $T = 1$. (c) and (d) Corresponding Floquet phase diagrams mapped as a function of the driving time period $T$, display the quantized values of $W_0$ and $W_{\pi}$ at a constant exchange coupling of $J_{sd} = 1$. The sharp boundaries between different color sectors explicitly denote phase transitions mediated by the closing and reopening of the bulk quasi-energy gaps. The remaining system parameters are the same as in Fig. \ref{fig:delk_spectrum}.}
\label{fig:ky_vs_jsd-tp_delk}
\end{figure}

Here, the +(-) sign corresponds to the emergence of FMZEMs (FMPEMs). This transcendental equation allows for an exact mapping of the bulk gap closing points in the $(T, \mu, J_{sd})$ parameter space at the high-symmetry points. We then numerically calculate the winding numbers as functions of the system parameters. Fig. \ref{fig:ky_vs_jsd-tp_delk} (a) and (b) displays the variation of these invariants with respect to $J_{sd}$, while Fig. \ref{fig:ky_vs_jsd-tp_delk} (c) and (d) illustrate the transition points as a function of the driving period $T$.

A striking feature of the delta-kicked phase diagrams is the reciprocal relationship between the zero and $\pi$ modes across the Brillouin zone, specifically $W_\pi (k_y+\pi) = -W_0 (k_y)$. The origin of this symmetry lies in the transformation properties of the symmetrized Floquet Hamiltonians, $H_{F1}$ and $H_{F2}$, under a momentum shift of $k_y \rightarrow k_y+\pi$.

We identify the unitary operators $U'= \sigma_z$ and $U''= \tau_z \sigma_z$ as the generators of these transformations for $H_{F1}$ and $H_{F2}$, respectively. By applying these to the integral in the winding number defined in Eq. (\ref{eq:winding_number}) and invoking the cyclic property of the trace, we can analyze the shift in the invariants. For the first symmetric frame, since $U'$ anti-commutes with the chiral symmetry operator $\Gamma$, the winding number picks up a sign change:

\begin{equation}
W_1(k_y + \pi) = -W_1(k_y).
\end{equation}

Conversely, for the second frame, the operator $U''$ commutes with $\Gamma$, leaving the invariant unchanged:

\begin{equation}
W_2(k_y + \pi) = W_2(k_y).
\end{equation}

Recalling the definitions of the Floquet winding numbers from Eq. (\ref{eq:floquet_winding_number}), 
we substitute the shifted values:
\begin{equation}
\begin{aligned}
W_\pi(k_y + \pi) &= \frac{W_1(k_y + \pi) - W_2(k_y + \pi)}{2} \\
&= \frac{-W_1(k_y) - W_2(k_y)}{2} \\
&= -W_0(k_y).
\end{aligned}
\end{equation}

This confirms that the emergence of $\pi$-modes at a given momentum is intrinsically linked to the zero-mode topology at the shifted momentum, providing a powerful analytical constraint on the global phase structure of the system.

\section{Extended Space Hamiltonian}
\label{extended_space}

To implement the Floquet Green’s function approach, we explicitly construct the extended Hamiltonian in Sambe space using the Fourier harmonics of the periodic drive. The infinite-dimensional matrix representation of the extended Floquet Hamiltonian $H^{\text{ext}}_F$ is given by:

\begin{equation}
    H^{\text{ext}}_F = 
    \begin{bmatrix}
    \ddots & \vdots & \vdots & \vdots & \vdots & \iddots \\
    \cdots & H_{(0)}-\Omega & H_{(-1)}  & H_{(-2)} & H_{(-3)}& \cdots \\
    \cdots &H_{(1)} & H_{(0)} & H_{(-1)} & H_{(-2)}& \cdots \\
    \cdots & H_{(2)} &H_{(1)} &  H_{(0)}+\Omega & H_{(-1)} & \cdots \\
    \cdots & H_{(3)} & H_{(2)} & H_{(1)} &  H_{(0)}+2\Omega& \cdots \\
    \iddots& \vdots & \vdots & \vdots & \vdots & \ddots
\end{bmatrix}
\end{equation}
where $\Omega = 2\pi/T$ is the fundamental driving frequency. The blocks $H_{(m)}$ are the Fourier components of the time-dependent Hamiltonian $H(t)$ which are defined as:

\begin{equation}
    H_{(m)}= \frac{1}{T} \int_0^T H(t) e^{im\Omega t} dt.
\end{equation}

For the square-wave protocol, the chemical potential is modulated as a piecewise constant function. The zeroth harmonic ($m=0$) corresponds to the time-averaged Hamiltonian:

\begin{equation}
H_{(m=0)} = H_0 - \bar{\mu} \tau_z,
\end{equation}
where $\bar{\mu} = (\mu_1 t_1 + \mu_2 t_2)/T$ is the effective static chemical potential. The off-diagonal blocks $H_{(m\neq 0)}$, which describe the absorption or emission of $\lvert m \rvert$ photons, are given by:

\begin{equation}
    H_{(m \neq 0)} = -\frac{i}{2m\pi}(\mu_1 - \mu_2)(1-e^{im\Omega t_1}) \tau_z.
\end{equation}

In the symmetric case where $t_1=t_2=T/2$ and $\mu_1 = -\mu_2=\mu$, the harmonics simplify to:

\begin{equation}
    H_{(m \neq 0)} =
    \begin{cases} 
      0 & \qquad m \quad\text{even}, \\
      -\frac{2i\mu}{m \pi} \tau_z & \qquad m \quad\text{odd}.
   \end{cases}
\end{equation}

Notably, these harmonics decay as $1/m$, implying that high-order multiphoton processes are progressively suppressed, which justifies the truncation of the Sambe space for numerical convergence.

In contrast, the Dirac delta-kick protocol presents a fundamentally different coupling structure. Due to the properties of the delta function, each kick contributes equally to all frequency harmonics:

\begin{equation}
    H_{(m)} =
    \begin{cases} 
      H_0 - \frac{\mu}{T} \tau_z & \qquad m = 0, \\
      -\frac{\mu}{T} \tau_z & \qquad m \neq 0.
   \end{cases}
\end{equation}

Unlike square-wave drive, the Fourier amplitudes of a delta-kick are independent of the harmonic index $m$. Consequently, all Floquet sidebands are coupled with equal strength, requiring a substantially larger Sambe-space truncation for numerical convergence and producing a richer quasienergy spectrum.

\bibliography{ref.bib}
\end{document}